\documentclass[twocolumn,epsf,prb,amsmath,amssymb,showpacs]{revtex4}
\usepackage{graphicx}
\usepackage{amsmath}
\usepackage{sidecap}
\usepackage{subfigure}

\begin{document}
\title{Snake states in graphene quantum dots in the presence of a p-n junction}

\author{M. Zarenia$^1$}
\author{J. M. Pereira Jr.$^2$}
\author{F. M. Peeters$^{1,2}$}
\author{G. A. Farias$^2$}
\address{$^1$Department of Physics, University of Antwerp,
Groenenborgerlaan 171, B-2020 Antwerpen, Belgium.\\
$^2$Departamento de F\'{\i}sica, Universidade Federal do
Cear\'a, Fortaleza, Cear\'a, 60455-760, Brazil.}

\date{ \today }

\begin{abstract}
We investigate the magnetic interface states of graphene quantum
dots that contain p-n junctions. Within a
tight-binding approach, we consider rectangular quantum dots in the presence of a perpendicular magnetic field
containing p-n, as well as p-n-p and n-p-n junctions. The results
show the interplay between the edge states associated with the
zigzag terminations of the sample and the snake states that arise at
the p-n junction, due to the overlap between electron and hole
states at the potential interface. Remarkable localized states are found at the crossing of the p-n junction with the zigzag edge having a dumb-bell shaped electron distribution. The results are presented as
function of the junction parameters and the applied magnetic flux.

\end{abstract}


\maketitle

\section{Introduction}
The study of graphene, a single layer of hexagonal carbon, has led
to the discovery of new phenomena that highlight the unusual
electronic properties of this 2D system \cite{Review}. In
particular, the linear gapless electronic spectrum, together with
the chirality of carriers in this system is predicted to allow 
perfect transmission through potential barriers (Klein paradox)
\cite{Katsnelson}. This transmission has a directional character and
is caused by the overlap between electron and hole states across the
potential barrier \cite{milton0,falko}. The effect has been
investigated experimentally in p-n junctions of gated graphene
samples \cite{Huard,Williams,Ozyilmaz,Stander}.

In the presence of an external magnetic field, the electron-hole
overlap at the potential barrier (or p-n junction) causes the
appearance of states that propagate along the junction interface
\cite{milton1,Marcus1}. These are known as {\it snake states} since,
in a semiclassical view, they arise through the coupling between
counter-circling cyclotron orbits on either side of the p-n
junction. They may also arise due to the presence of inhomogeneous
fields\cite{orozlany} and in warped and folded
graphene\cite{prada,rainis}. The presence of snake states was found
to influence the electronic properties of graphene-based samples in
the quantum Hall regime \cite{Marcus1}. Moreover, the coupling of
snake states have been predicted to modify electrical current
transport near the interfaces of narrow p-n-p junctions
\cite{Levitov}. Recently, experimental evidence was provided of the chaotic
coupling of snake states in quantum point contacts\cite{Marcus2}.
\begin{figure}[]
\centering
   \includegraphics[width=7.5 cm] {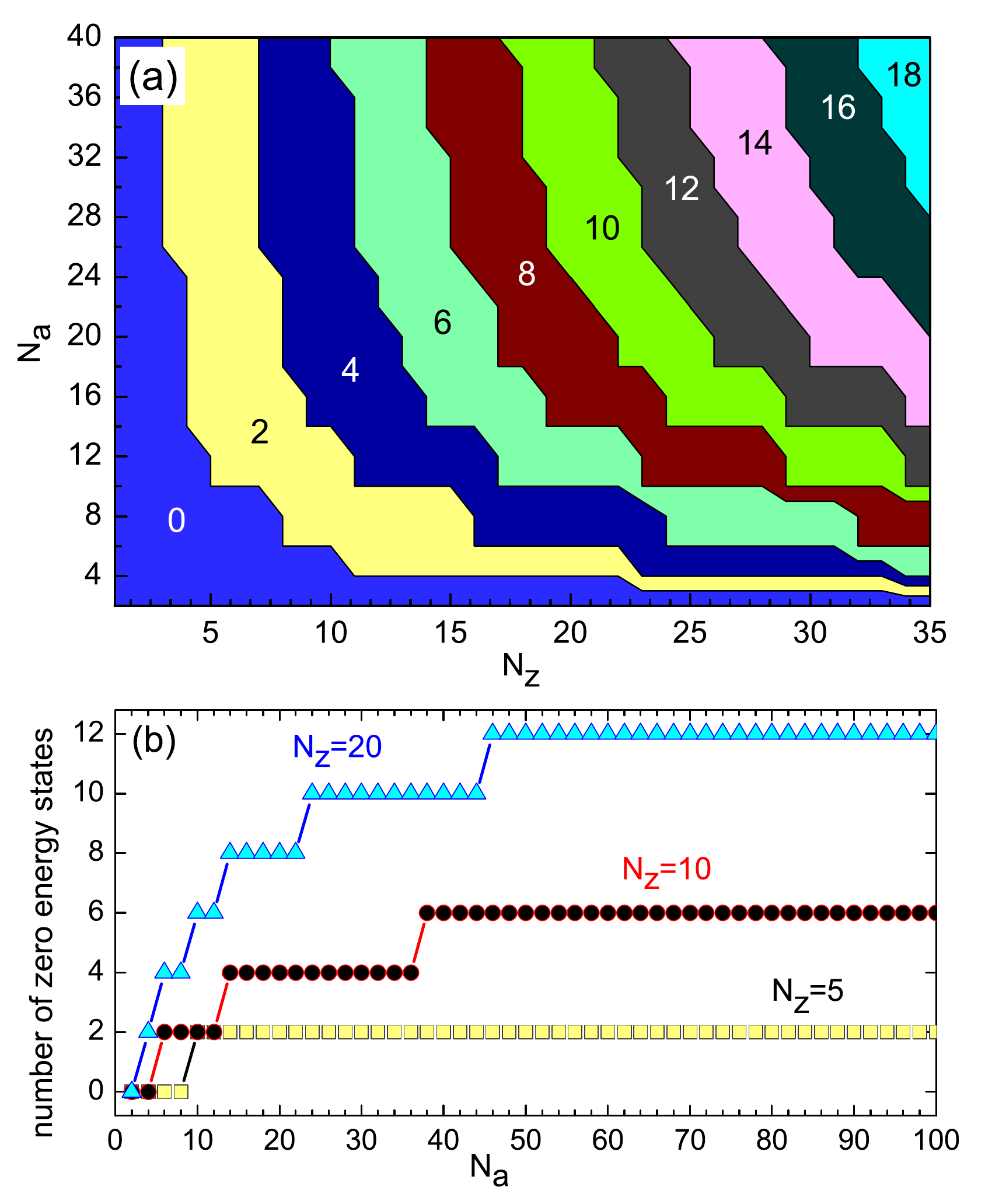}
\caption{(Color online) (a) Number of nearly zero energy states within the
interval $|E|<0.1$ meV as function of armchair ($N_a$) and zigzag
($N_z$) edge atoms. (b) Number of zero energies as function of $N_a$
for $N_z=5,10,20$.}\label{fig1}
\end{figure}
\begin{figure}[]
\centering
   \includegraphics[width=8. cm] {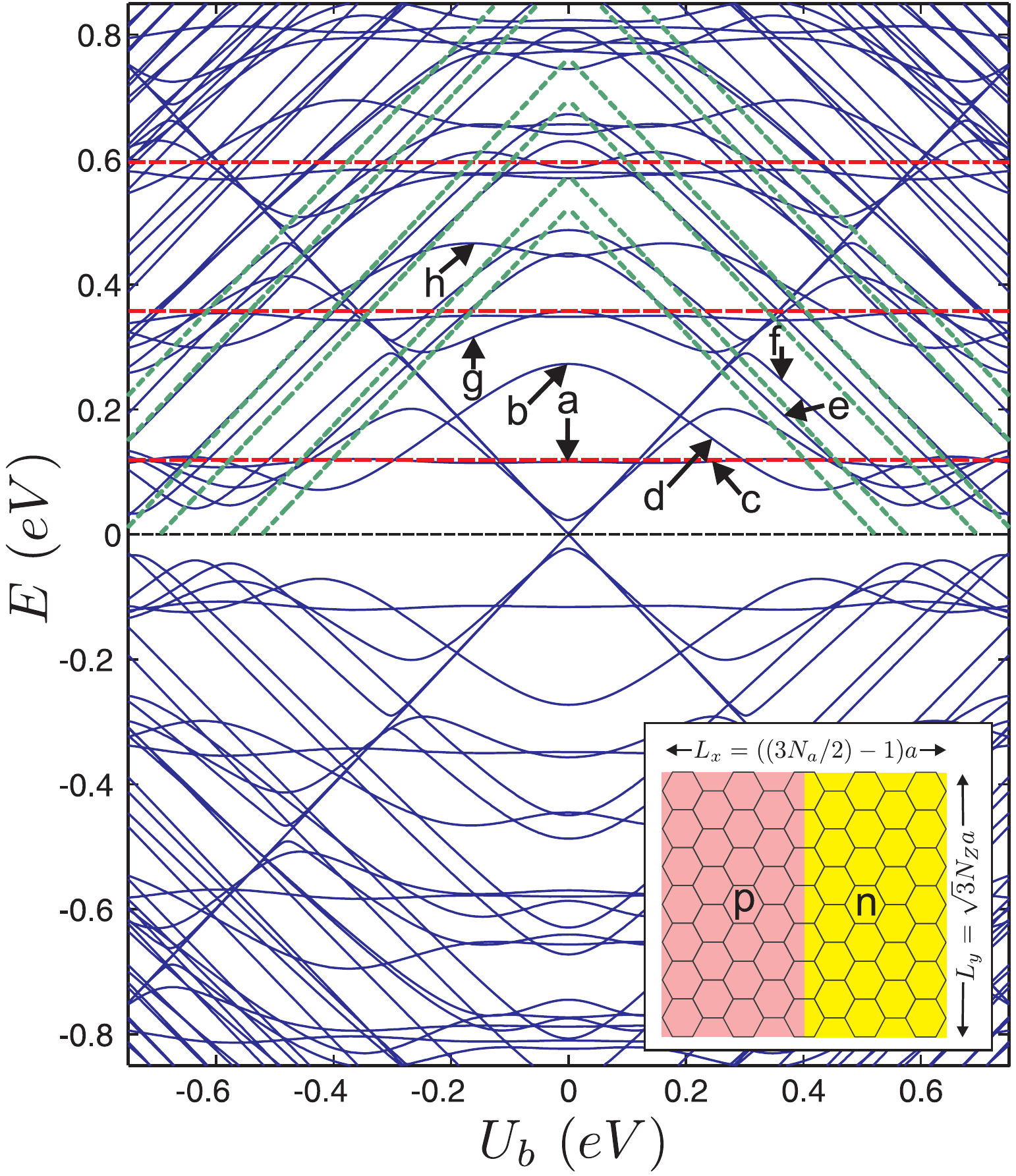}
\caption{(Color online) Energy levels of a system with a p-n junction parallel to
the zigzag edges in a rectangular GQD with $L_{x}=7.52$ nm,
$L_y=8.6$ nm and $\Phi_c/\Phi_{0}=0$ as function of applied voltage
$U_b$. The system is illustrated in the inset. The different colors
represent the different gate voltages ($+U_{b}$ for p-type and
$-U_{b}$ for n-type). The red dashed lines correspond to the energy
states given by Eq. (3) and the green dashed lines represent the
energy levels obtained from Eq. (4).}\label{fig2}
\end{figure}

In this paper we investigate theoretically the interplay between
edge and snake states of p-n, p-n-p, and n-p-n junctions imposed on
graphene quantum dots (GQDs). We study the character of the different confined states by looking at the probability densities. The electron probability density can be linked to the local density of states (LDOS) which is a quantity that can be measured experimentally using scanning tunneling microscopy (STM). Measurement of the LDOS allows the probing of the spatial structure of the confined energy levels. Such measurements were recently reported for graphene quantum dots\cite{hmalainen}. We consider GQDs created by cutting a larger graphene sample in order to obtain electronic confinement in
a nanometer-scale structure with well-defined edges. The properties
of the confined states of such GQDs in a magnetic field have been
studied theoretically \cite{marko,Ensslin0} as well as
experimentally \cite{Ensslin2}. Note that such p-n junctions (but of
irregular shape) are also naturally present in graphene samples when
the Fermi energy is located around the Dirac point. They are
generally known as puddles and have investigated with scanning
tunneling microscopy (STM)\cite{martin,deshpande,zhangNa}. In our calculations we neglect disorder which for the considered small sized dots will be of secondary importance.
\begin{figure}[!]
\centering\subfigure{
   \includegraphics[width=8cm] {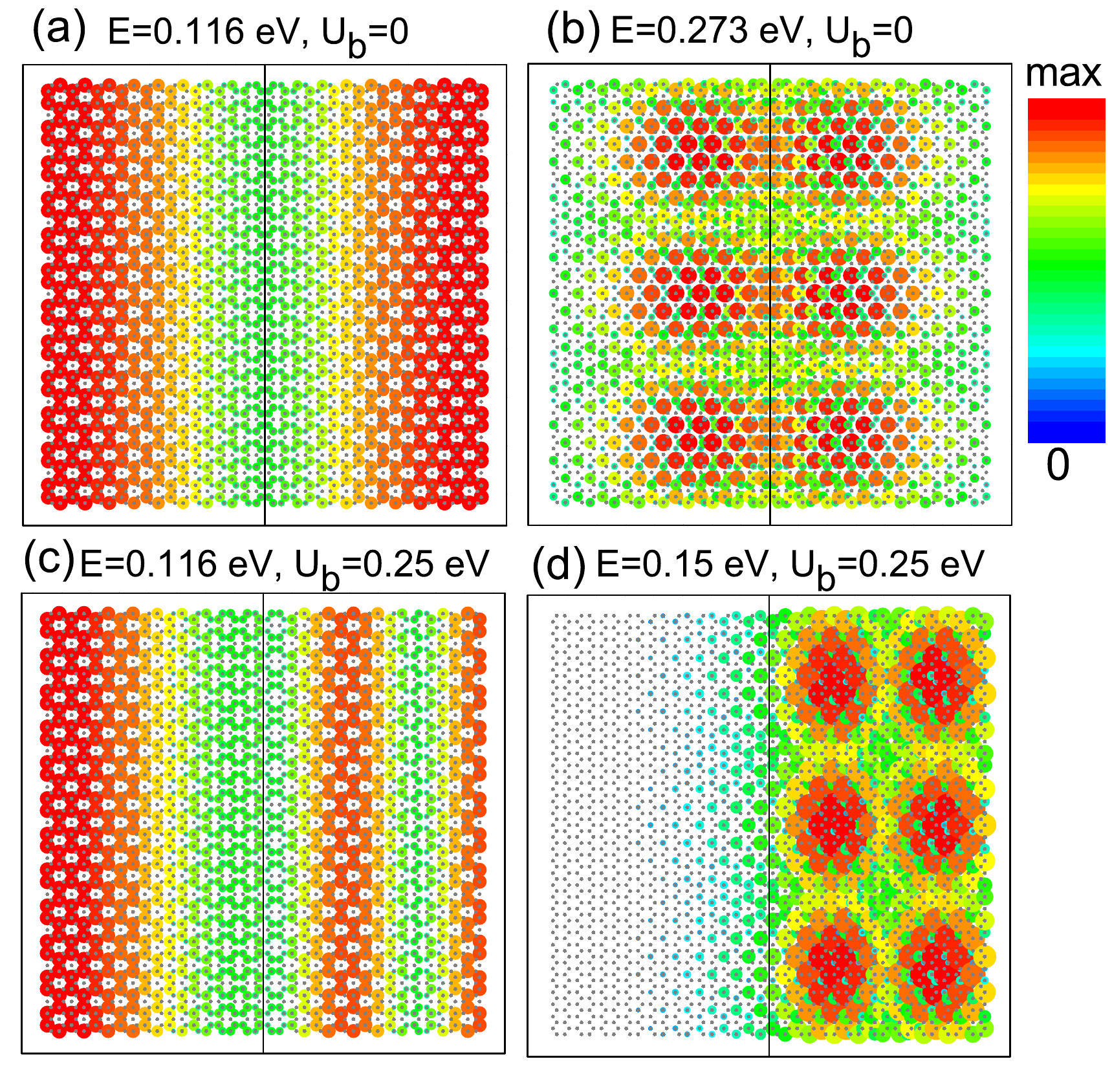}}
   \subfigure{
   \hspace{-1cm}
   \includegraphics[width=7.3cm] {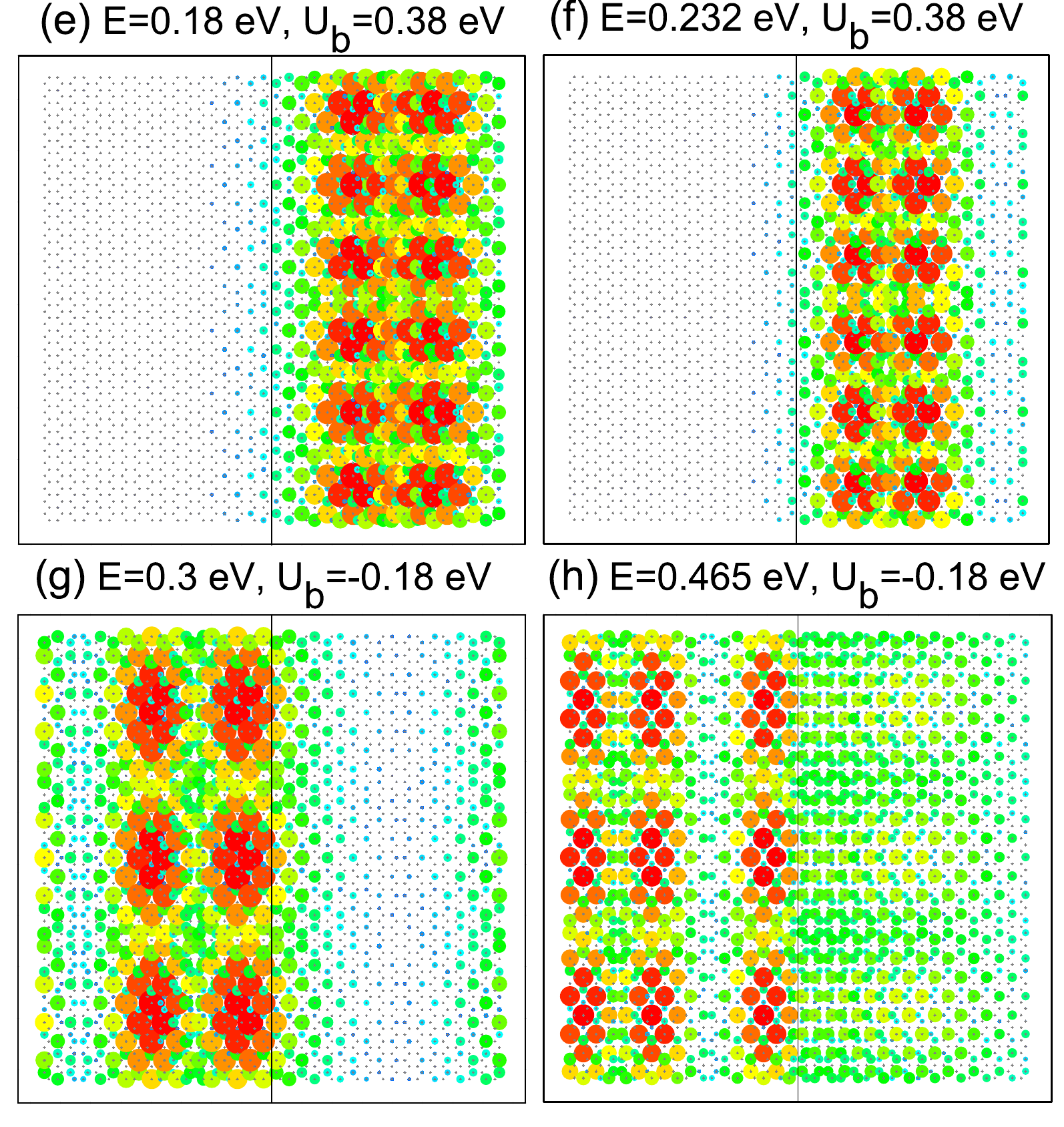}}
   \caption{(Color online) Probability densities corresponding to the points indicated
by (a-h) in Fig. \ref{fig2}. The black vertical line indicates the
position of the p-n junction.}\label{fig3}
\end{figure}

One important aspect of such graphene-based structures is the
possible existence of edge states, for which the wavefunctions are
localized at zigzag terminations of the sample
\cite{Nakada,Kohomoto,Tang,Zhang,zarenia}. These states have been
recently observed by STM\cite{Kobayashi,Niimi}. The presence of edge
states can be especially relevant for nanometer-scale graphene
structures. In particular, depending on the geometry of the GQDs,
the edge states can correspond to the ground state of the system
\cite{Zhang}. For GQDs of general shape, Wimmer {\it et al.} have
shown that the edge states tend to form a narrow band and are
generally robust with regards to perturbations \cite{Guinea}. In the
present case we consider the effect of a position-dependent
potential profile and an external perpendicular magnetic field on
the energy spectrum of rectangular GQDs in the context of the
nearest-neighbor tight-binding model. The presence of the potential
interface thus introduces additional localized states, i.e. snake
states, which can hybridize with the
conventional zigzag edge states. 

The paper is organized as follows: Section II gives a description of
the model. In Sec. III we show and discuss the analytical and
numerical results. Our summary and conclusions are presented in Sec.
IV.
\begin{figure}[!]
\centering
   \includegraphics[width=8. cm]{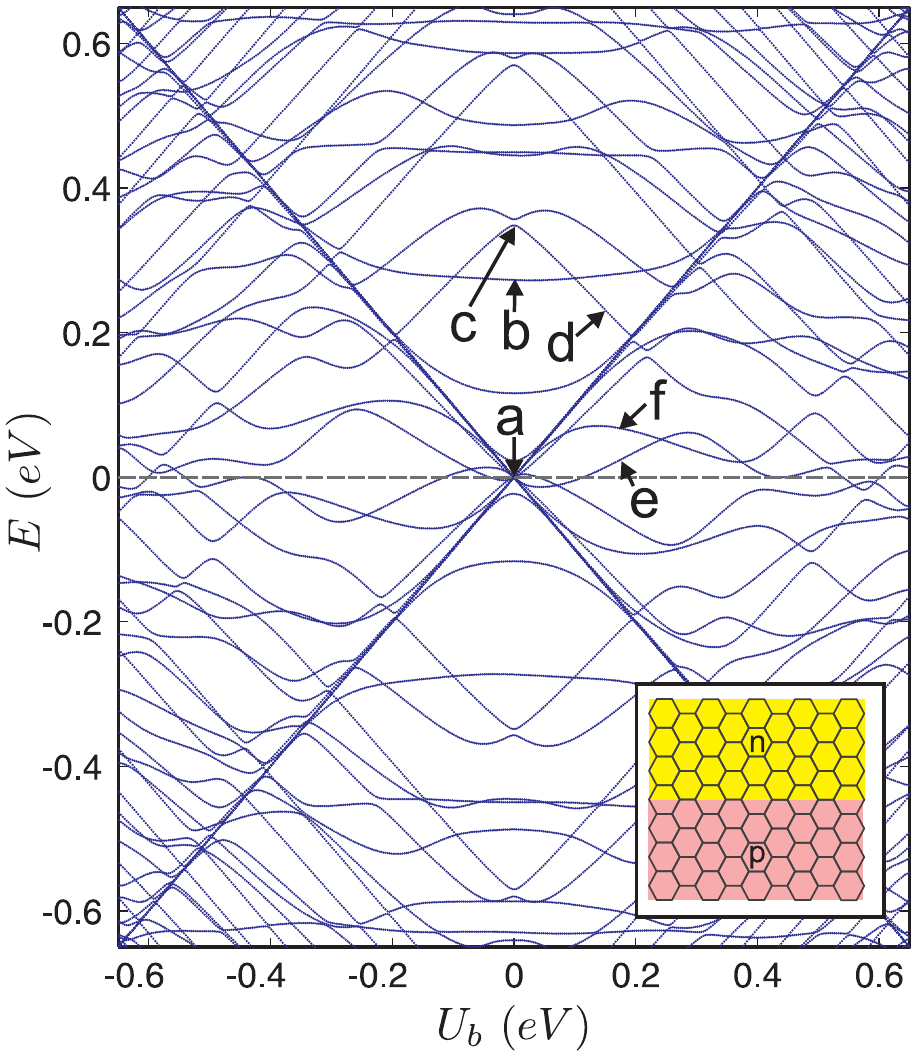}
\caption{(Color online) Energy levels of a rectangular GQD with a p-n junction
parallel to the armchair edges (see the inset of the figure) as
function of applied electrostatic potential $U_b$ for the same
parameters as Fig. \ref{fig2}.}\label{fig4}
\end{figure}
\begin{figure}[!]
\centering
   \includegraphics[width=8.5 cm] {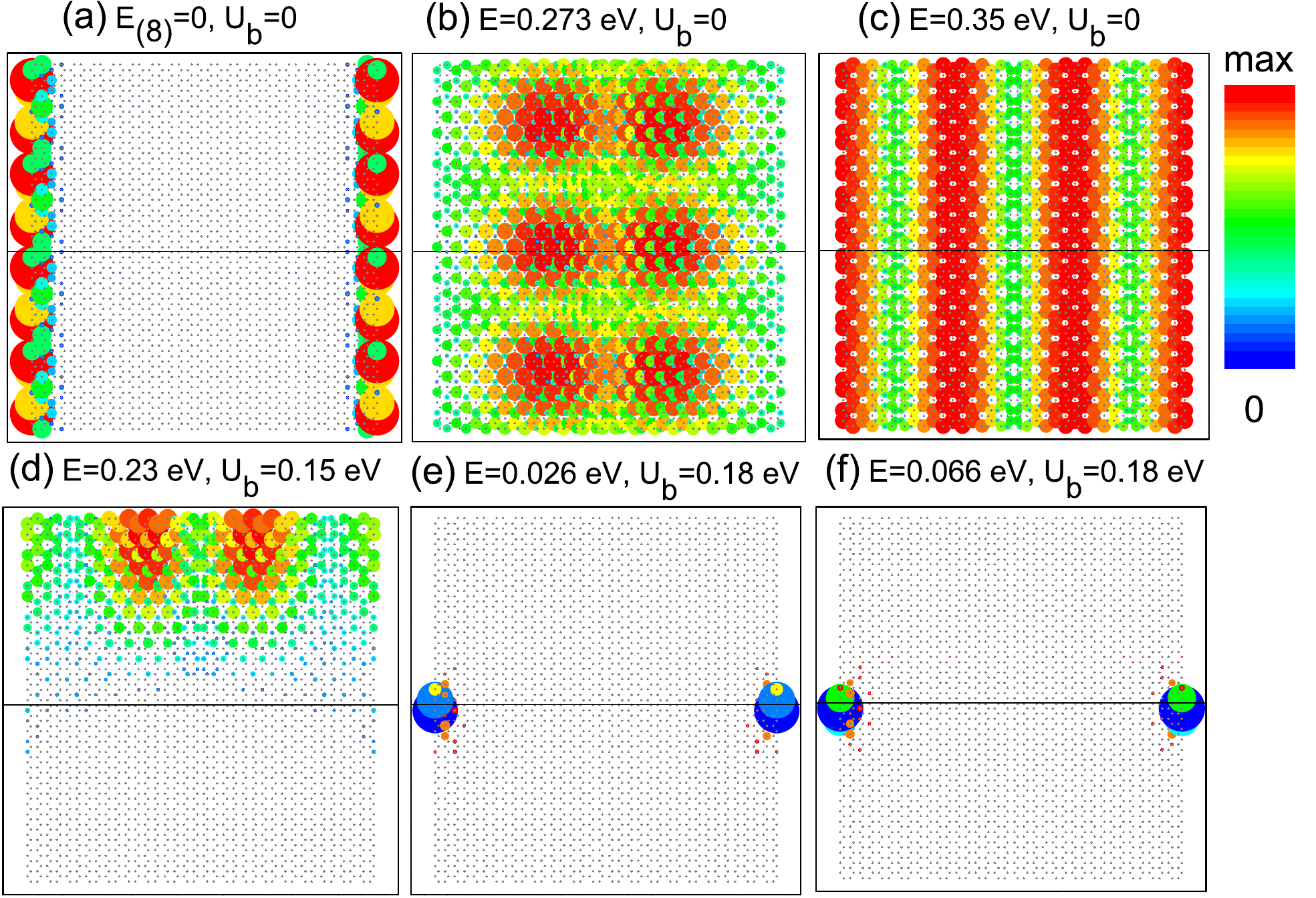}
\caption{(Color online) Probability densities corresponding to the
points indicated by (a-f) in Fig. \ref{fig4}. In panel (a) $E_{(8)}=0$ indicates
the eighth degenerate zero energy. The black horizontal
line indicates the position of the p-n junction.}\label{fig5}
\end{figure}
\section{Model}
The nearest-neighbor tight-binding Hamiltonian of the $\pi$
electrons in the honeycomb graphene lattice can be written as
\begin{equation}
{\mathcal H}= \sum_m \epsilon_m c_m^{\dagger}c_m + \sum_{lm}
(t_{lm} c_l^{\dagger}c_m + h.c.),
\end{equation}
where $\epsilon_m$ is the on-site energy, $t_{lm}$ is the
nearest-neighbor coupling parameter and $c_j$ ($c_j^{\dagger}$) is
the annihilation (creation) operator of the electron at a site with
label $j$. The external magnetic field introduces the Peierls phase
in the coupling term $t_{lm} = t\exp\big(2\pi i\Phi_{l,m}\big)$,
where $t$ is the zero-magnetic field coupling parameter, $\Phi_{l,m}
= (1/\Phi_0)\int_{r_l}^{r_m}{\mathbf A}\cdot {\mathbf dr}$, and
$\Phi_0=h/e$ is the magnetic quantum flux and ${\mathbf A}$ is the
vector potential. For graphene one has $t = 2.7$ eV. The field is
given by ${\mathbf B} = B$\^{z} and we choose the Landau gauge as 
${\mathbf A}= (0,Bx,0)$. Then, the Peierls phase for a transition
between two sites $l$ and $m$ is $\Phi_{l,m} = 0$ in the $x$
direction and $\Phi_{l,m} = \pm(x/3a)\Phi_c/\Phi_0$ along the $\pm
y$ direction, where $\Phi_c=3\sqrt{3}a^{2}B/2$ is the magnetic flux
threading one carbon hexagon with $a=0.142$ nm being the C-C
distance. The p-n, p-n-p or n-p-n junctions are modeled by assuming
a position-dependent on-site energy $\epsilon_m = \epsilon(m)$.
Throughout this paper we assign the values $\epsilon = U_b$
($\epsilon = -U_b$) for the p (n) regions, whereas at the interfaces
between these regions the potential is assumed to vary abruptly. This assumption is expected
not to influence the results qualitatively.
\section{p-n junction: zero magnetic field}
We consider an almost square quantum dot because it allows us to
investigate the effect of both armchair and zigzag edges in the same
sample. We are interested to learn how the confined states are
influenced by the relative orientation of the p-n interface with
respect to the specific type of edges. Here, the length of the
rectangular GQD which is terminated by armchair
edges is defined as $L_{x}=[(3N_a/2)-1]a$ and the length terminated
at the zigzag edges is $L_{y}=N_{z}\sqrt{3}a$ where $N_{a}$ and
$N_z$ are the number of C-atoms, respectively at the armchair and
zigzag edges. The total number of C-atoms in the rectangular GQD is
$N=N_{a}(2N_{z}+1)$. We should notice that the energy spectrum of a
rectangular GQD exhibits zero energy states\cite{Kim} which are
confined at the zigzag edges. The number of zero energy states in a
rectangular dot depends to the number of both armchair and zigzag
atoms\cite{Kim}. Figure \ref{fig1}(a) shows the number of states
with \emph{nearly} zero energies (we took the number of states
within the energy interval $|E|<0.1$ meV) as function of $N_z$
and $N_a$. Our results show that for a fixed number of zigzag edge
atoms the number of zero-energy states increases with increasing the
armchair edge atoms (see Fig. \ref{fig1}(b)). Notice that the number
of zero-energy states can not exceed $2N_z$.
\begin{figure}[!]
\centering \subfigure{
\hspace{-1cm}
   \includegraphics[width=8 cm] {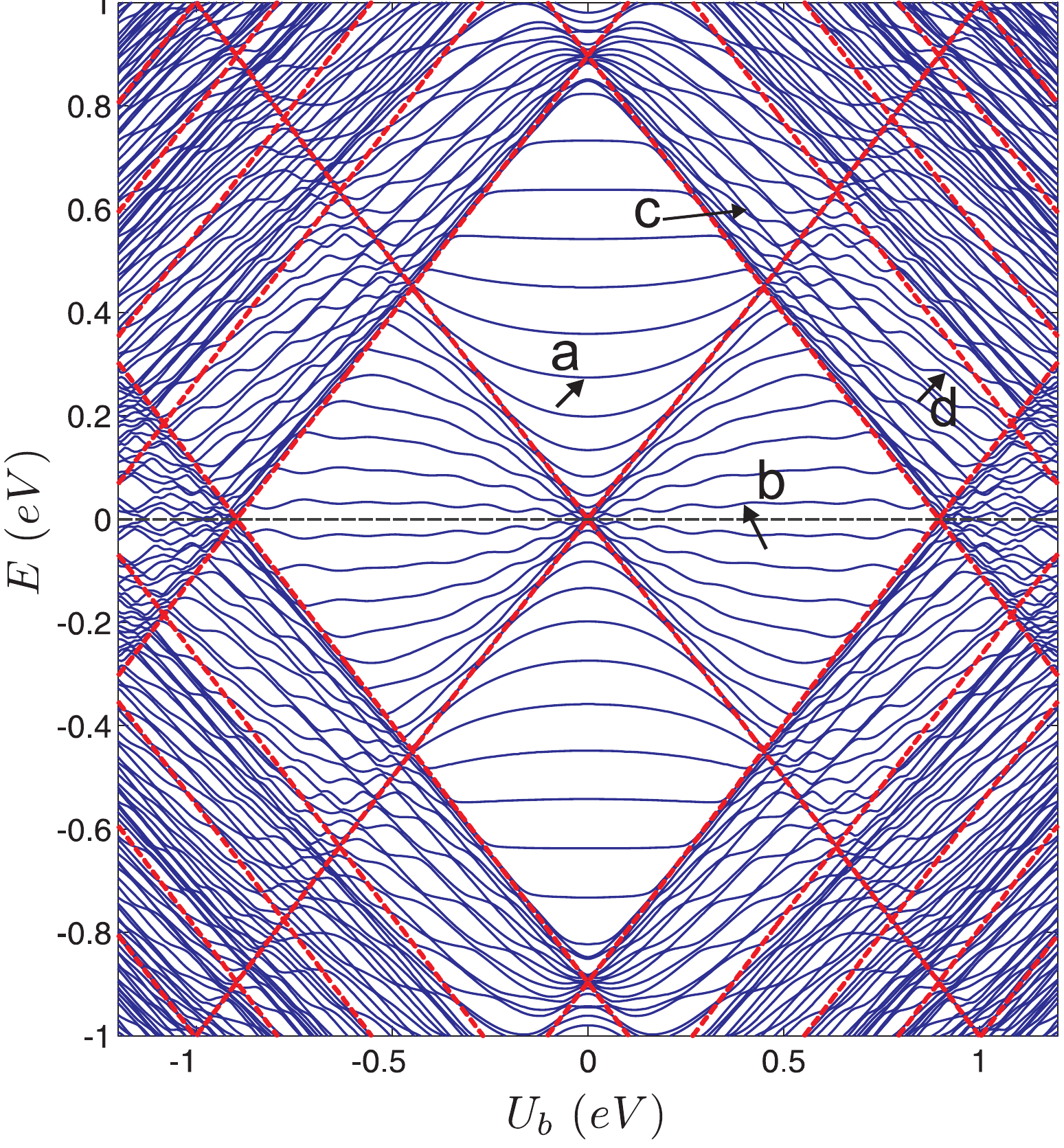}
    }
  \subfigure{
    \includegraphics[width=8cm] {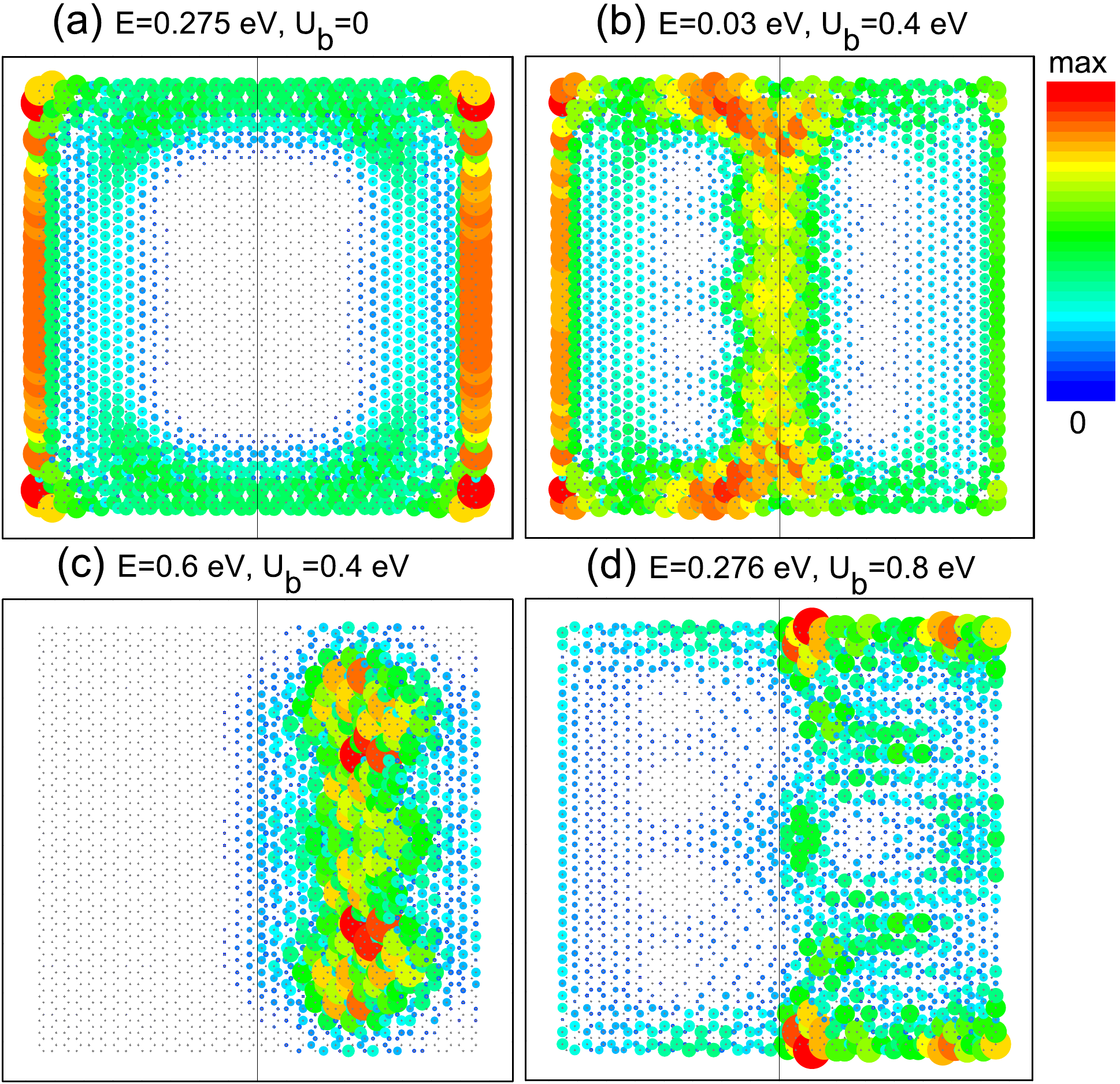}
   }
\caption{(Color online) Upper panel: The same as Fig. \ref{fig2} but
with non-zero magnetic flux $\Phi_c/\Phi_{0}=0.1$. The red dashed
lines are the LLs of an infinite graphene sheet which are shifted up
(down) by $U_b$($-U_b$). Lower panels: Probability densities
corresponding to the points indicated by (a-d) in the upper panel.
The black vertical line indicates the position of the p-n
junction.}\label{fig6}
\end{figure}
\begin{SCfigure*}
\hspace{1.5cm}
\includegraphics[width=10 cm]{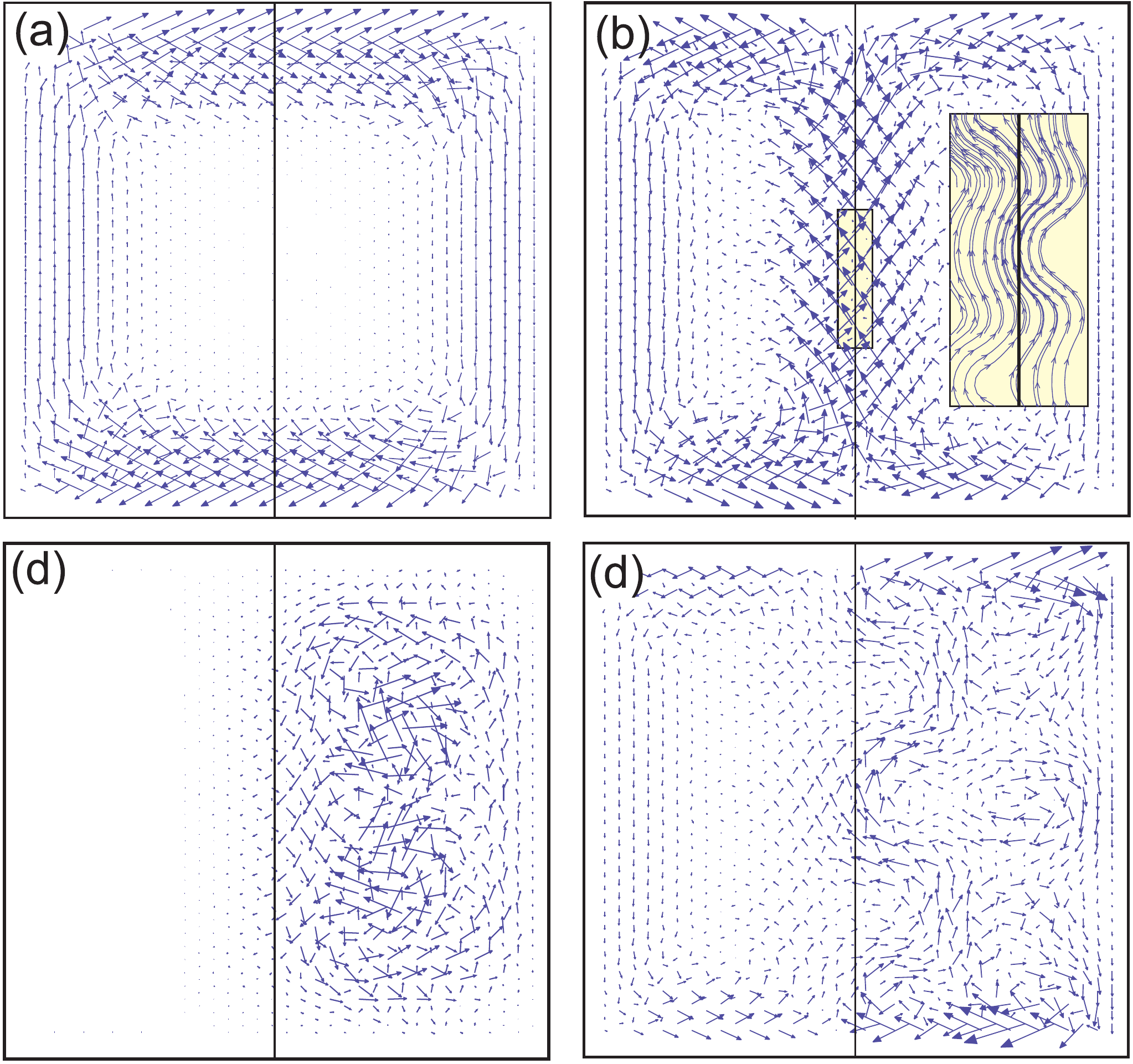}
\caption{(Color online) Current density profile corresponding to the
states indicated by (a-d) in Fig. \ref{fig6}. The black line
indicates the position of the p-n junction. The inset in (b) shows a streamline plot of the enlargement region.}
\label{fig7}
\end{SCfigure*}

Now we solve the Hamiltonian (1) for a system in which a p-n
junction is parallel to the zigzag edges of a rectangular GQD (see
the inset of Fig. \ref{fig2} where different colors represent the
p-type and n-type regions which are respectively subjected to
$+U_{b}$ and $-U_b$ gate voltages). For numerical purposes we take
as an example $N_a=36$ ($L_x=7.52$ nm) and $N_z=35$ ($L_{y}=8.6$ nm)
in all the results of this paper. The energy levels of this system
are shown as function of the gate voltage $U_b$ in Fig. \ref{fig2}
for zero magnetic flux $\Phi_{c}/\Phi_{0}=0$. In the presence of a
p-n junction parallel to the zigzag edges the zero energy-degenerate
states split into two groups of degenerate states with energy:
\emph{i)} $E=+U_{b}$ and \emph{ii)} $E=-U_b$ where the number of the
states in each group is equal. Note that the energy spectrum in Fig.
\ref{fig2} exhibits a group of states which on average are almost independent
of $U_{b}$. Figure \ref{fig3} shows probability density plots for
the states indicated by the letters (a-h) in Fig. \ref{fig2}. The
probability densities in Fig. \ref{fig3}(a) and Fig. \ref{fig3}(c)
exhibit a nodal character across the zigzag edges and consequently
the $U_b$ energy shift from the p- and n-regions cancel out.
These levels are similar to confined states in a zigzag
\emph{nanoribbon}. The energy levels of a zigzag nanoribbon are
described, using the continuum model, by the transcendental
equation\cite{brey}
\begin{figure}[!]
\centering\vspace{1.2 cm} \hspace{-0.3cm}
\includegraphics[width=7.5 cm] {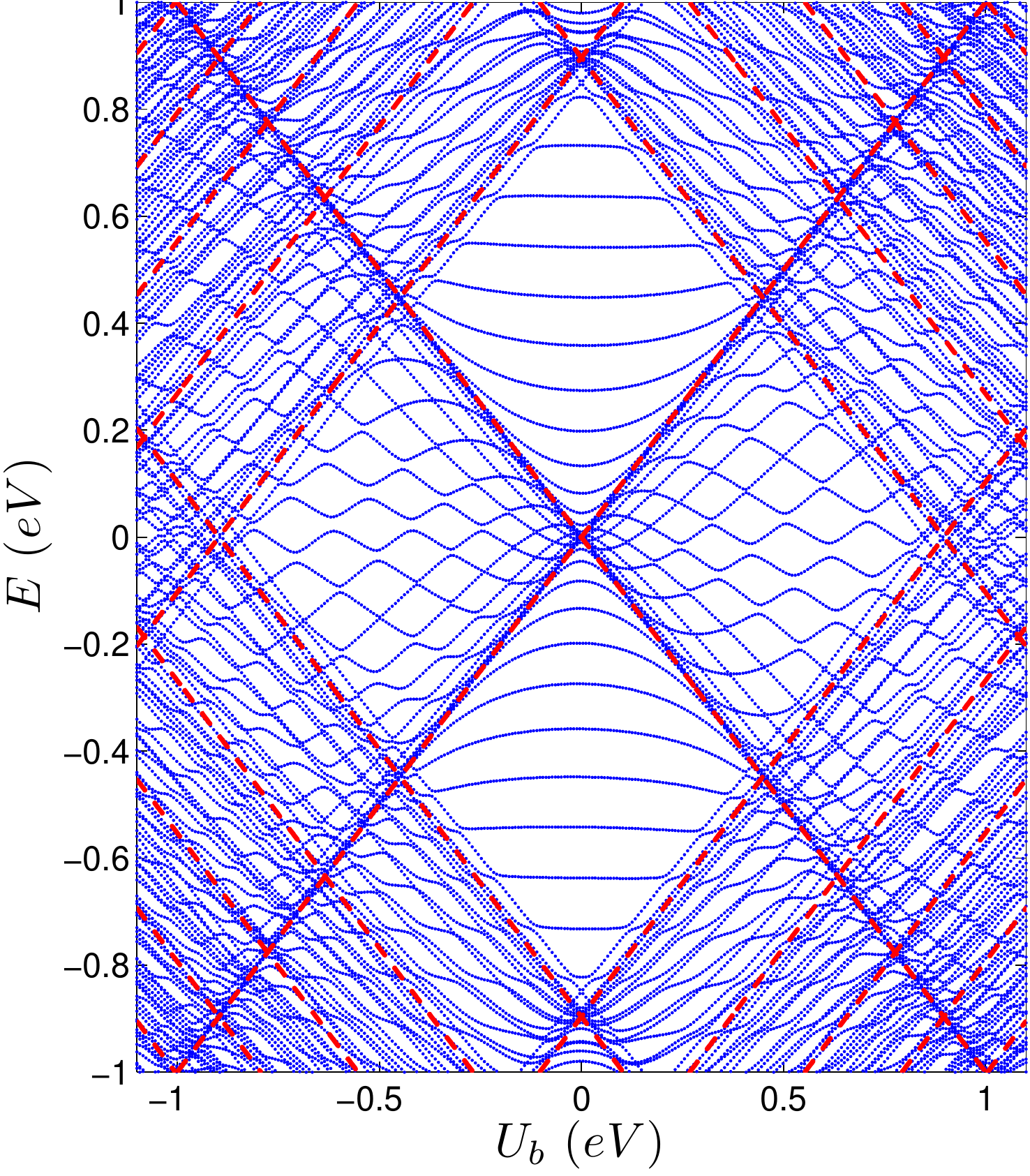}
\caption{(Color online) The same as Fig. \ref{fig4} but in the presence of magnetic flux
$\Phi_c/\Phi_{0}=0.1$. The red dashed lines are the LLs of an
infinite graphene sheet which are shifted up (down) by $U_b=0.25$ eV
($-U_b$).}\label{fig8}
\end{figure}
\begin{equation}
    \frac{k_y-\kappa}{k_y+\kappa}=e^{-2L_{x}\kappa}
\end{equation}\label{zig}\noindent
where $\kappa=\sqrt{k_{y}^2-(\epsilon/\hbar v_{F})^2}$ with $v_{F}=10^6$ m/s being the Fermi velocity. In the low energy limit we take $k_{y}=0$ where Eq. (2) becomes $\displaystyle{\exp(\pm2iL_x\epsilon/\hbar v_{F})}=-1$ and results in
\begin{equation}
\epsilon_n=\pm\frac{\pi\hbar
v_{F}}{L_{x}}\big(n+\frac{1}{2}\big),~~~~n=0,1,2,...
\end{equation}\label{eq}\noindent
The three first electronic levels of the above relation are shown in
Fig. \ref{fig2} by red dashed lines which coincide reasonable well
with the position of the constant energy levels in the spectrum.
Note that the agreement is better for low energy where the continuum
model is more accurate. For the levels where the wavefunction is
spread out inside the dot the energy levels are approximately linear
with $U_{b}$. As seen in Figs. \ref{fig3}(b,d) the probability
density corresponding to these levels shows an oscillatory behavior
along the y-direction which is due to the confinement by the armchair edges. For
armchair nanoribbons the wave vector $k_{y}$ satisfies the condition
$k_{y}=(n_y\pi/L_y)+(2\pi/3\sqrt{3}a)$ where $n_y$ being an
integer\cite{brey}. Using Eq. (3) we take $k_x=\pi(n_x+1/2)/L_x$
along the x-direction. Thus the corresponding energies in the
presence of $\pm U_b$ are proportional to $\pm(U_b\pm\hbar
v_F\sqrt{k_y^2+k_x^2})$ which results in
\begin{equation}
\epsilon_{n}=\pm\left[U_{b}\pm\hbar
v_{F}\sqrt{\big(\frac{n_y\pi}{L_{y}}+
\frac{2\pi}{3\sqrt{3}a}\big)^2+\big(\frac{\pi(n_x+1/2)}{L_x}\big)^2}\right]
\end{equation}\label{arm}
These electronic levels described by Eq. (4)
which are shown by the green
dashed lines in Fig. \ref{fig2} for $-26\leq n_y\leq-20$ and $n_x=1$. The above arguments describe
reasonably well qualitatively most of the energy levels that are
found in the numerical spectrum depicted in Fig. \ref{fig2}. Because
of the finite boundaries those levels may interact leading to
anti-crossings. Aside from anti-crossings, the lines describe rather
well the low energy levels in the spectrum that decrease linearly
with $|U_b|$. Figs. \ref{fig3}(e,f) show the electronic density
corresponding to the lowest paired levels for $U_b=0.38$ eV where
the electrons are only confined in the $n$ region. Figures \ref{fig3}(g,h) show those states
that are influenced by both zigzag and armchair edges.
\begin{figure}[!]
\centering
\includegraphics[width=8 cm]{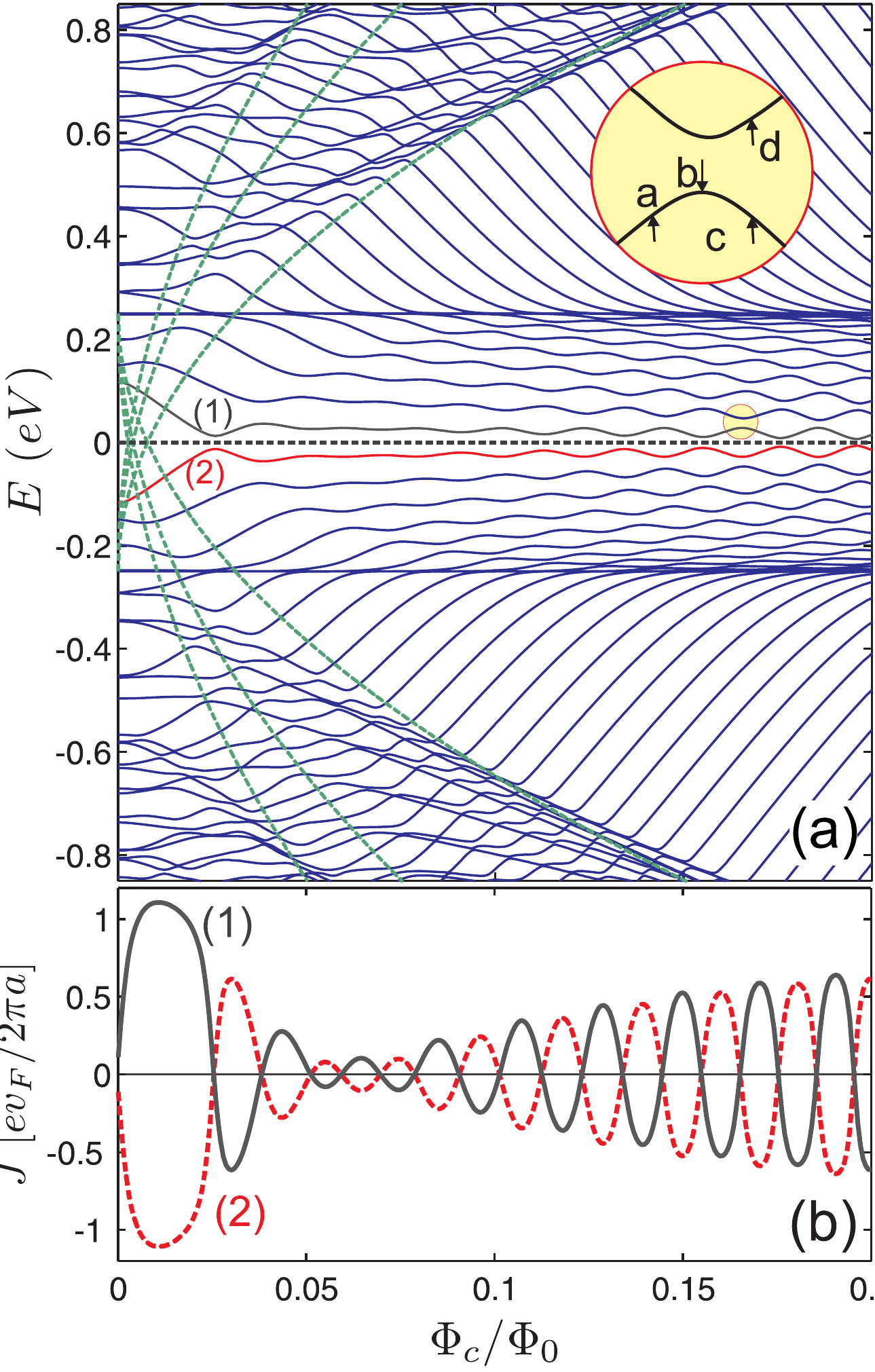}
\caption{(Color online) (a) Energy levels of a rectangular GQD subjected to a p-n
junction parallel to the zigzag edges (see the inset of Fig.
\ref{fig2}) as function of magnetic flux threading one carbon
hexagon $\Phi_{c}$ for $U_{b}=0.25$ eV. The green dashed curves are
the Landau levels of an infinite graphene sheet which are shifted up
(down) by $U_b$ ($-U_b$). (b) Persistent current corresponding to
the first electron state (gray solid curve, labeled by (1)) and the
first hole state (red dashed curve, labeled by (2)) as function of
external magnetic flux $\Phi_{c}$.} \label{fig9}
\end{figure}
\begin{figure}[!]
\centering
\includegraphics[width=7.5 cm]{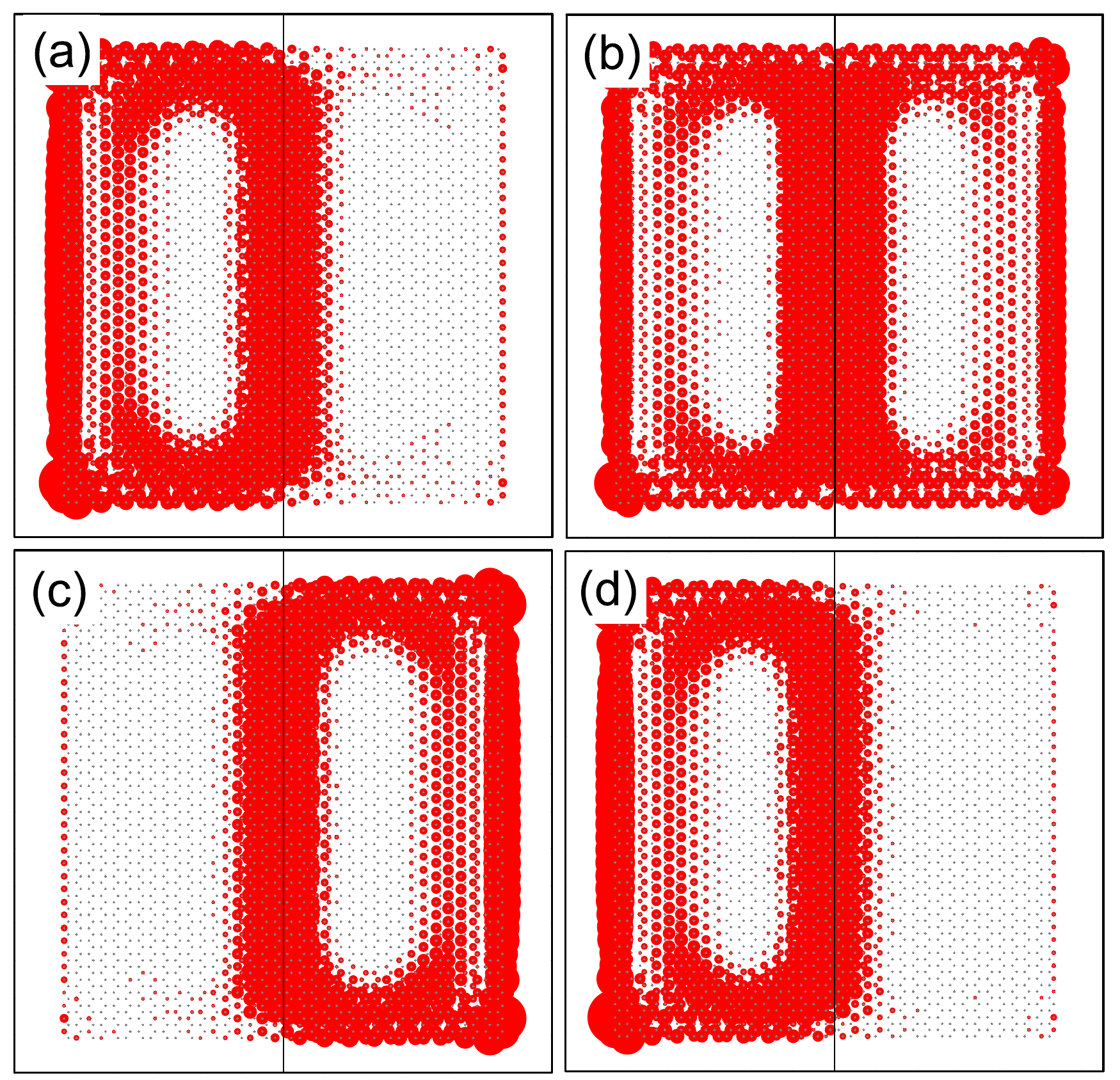}
\caption{(Color online) Probability densities corresponding to the points indicated
by (a-d) in the enlarged circle in Fig. \ref{fig9}(a). The black
line indicates the position of the p-n junction.} \label{fig10}
\end{figure}

The energy levels of a rectangular GQD subjected to a p-n junction
parallel to the armchair edges is shown in Fig. \ref{fig4}. The
system is depicted in the inset of Fig. \ref{fig4}. Since the p-n
interface is now located perpendicular to the direction of the edge
states (i.e. zigzag edges) the energy spectrum exhibits a complex
behavior as function of $U_{b}$. In this case the spectrum is not symmetric under switching $U_b \rightarrow -U_b$ for $|E|<U_b$ which is due to the fact that the number of p-type and n-type atoms are unequal. The probability density
corresponding to the points indicated by arrows are shown in Fig.
\ref{fig5}. For $U_{b}=0$ and $E=0$ the carriers are confined at the
zigzag edges (see Fig. \ref{fig5}(a) and the level is eighteen fold degenerate). Notice that the rectangular GQD
with $N_a=36$ and $N_z=35$ has 16 zero energy states (see Fig.
\ref{fig1}(a)). The probability density corresponding to the upper
states are spread out over the dot in both x- and y-directions (see Fig.
\ref{fig5}(b)) or along the zigzag edges (see Fig. \ref{fig5}(c)).
In the presence of a p-n junction the electrons confine at the p(n)-region when $E>U_b$ and the energy state decrease(increase) with $U_b$ (Fig. \ref{fig5}(d)). In contrast with Fig. \ref{fig2}, for a p-n junction parallel to the armchair edge several states are found  for $E < U_b$. These states, as seen in Figs. \ref{fig5}(e,f), present an interesting behavior: the probability densities show a significant localization at the intersection of the p-n interface and the zigzag edges. That can be explained as resulting from the hybridization of the zigzag edge states on each side of the p-n junction. This remarkable state appears only when the p-n junction crosses a zigzag edge. Note that the wavefunction of this state consists of both electron and hole components. 
%
\begin{figure}[!]
\centering
\includegraphics[width=8 cm]{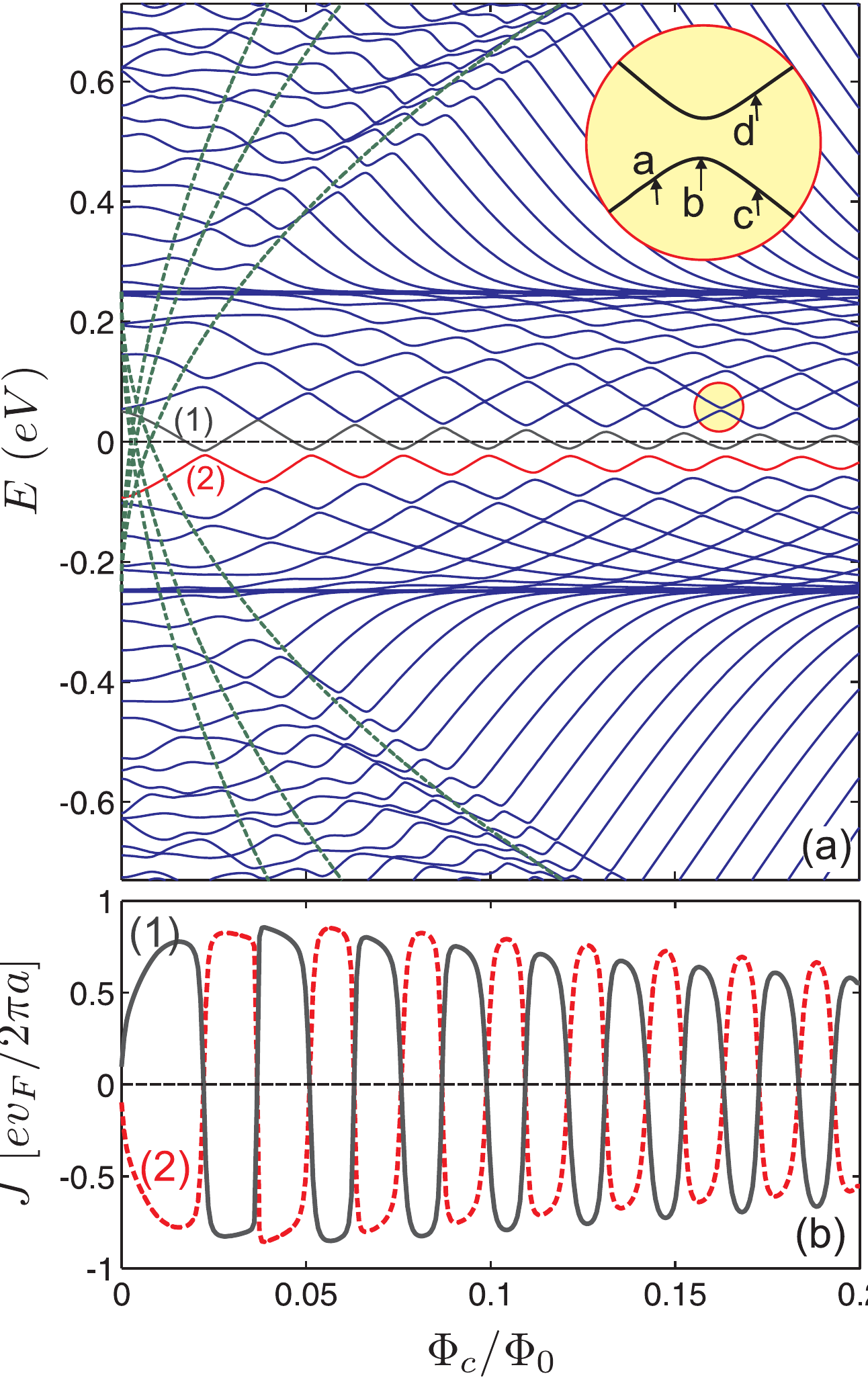}
\caption{(Color online) The same as Fig. \ref{fig9} but for a rectangular GQD
subjected to a p-n junction parallel to the armchair edges (see the
inset of Fig. \ref{fig4})} \label{fig11}
\end{figure}
\begin{figure}[!]
\centering
\includegraphics[width=8.2 cm] {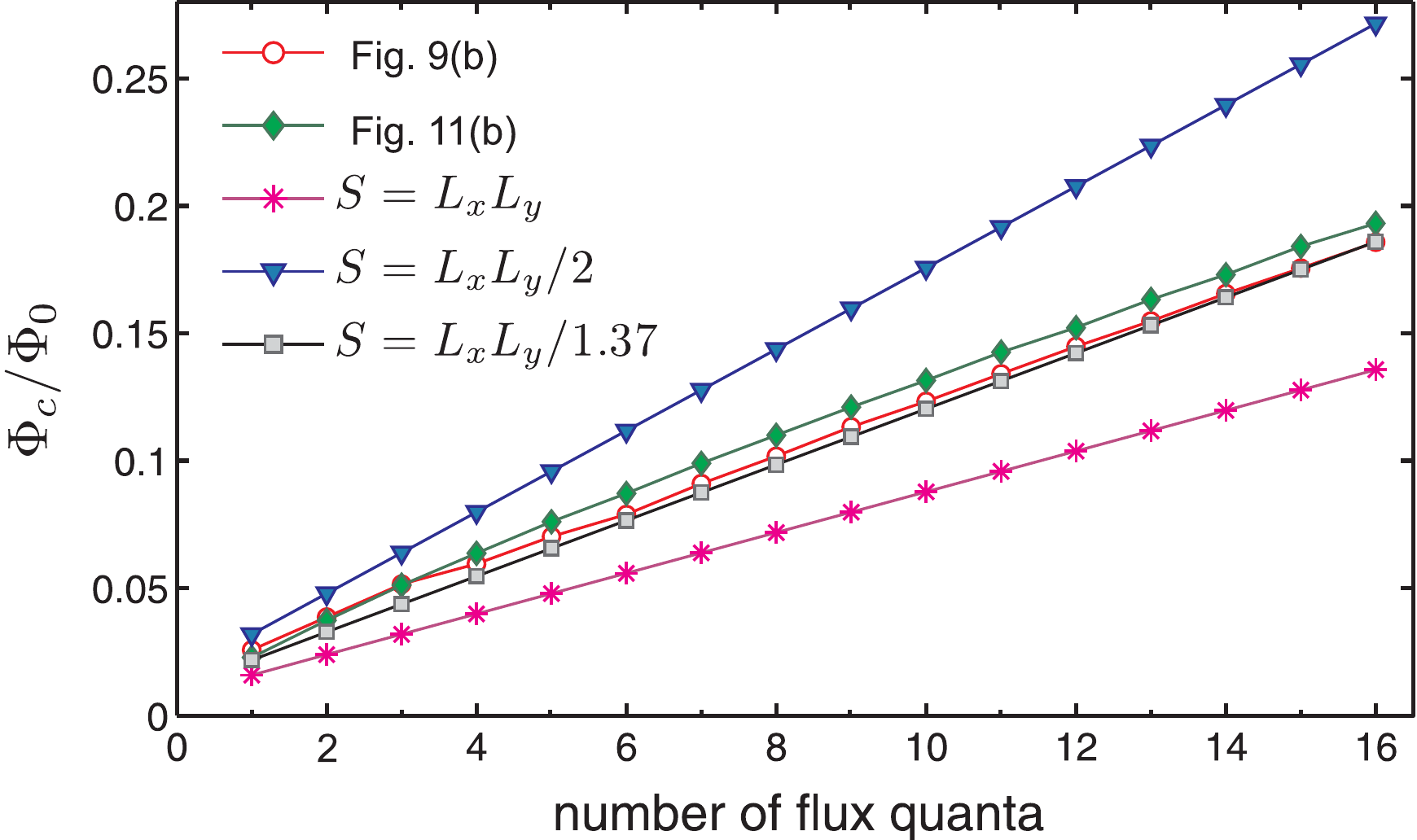}
\caption{(Color online) The position of the oscillations in Figs. \ref{fig9}(b) and \ref{fig11}(b) and the number of flux quanta through the surface area $S=L_xL_y, L_xL_y/2, S=L_xL_y/1.37$ as function of magnetic flux.}\label{fig12}
\end{figure}
\begin{figure}[!]
\includegraphics[width=8 cm]{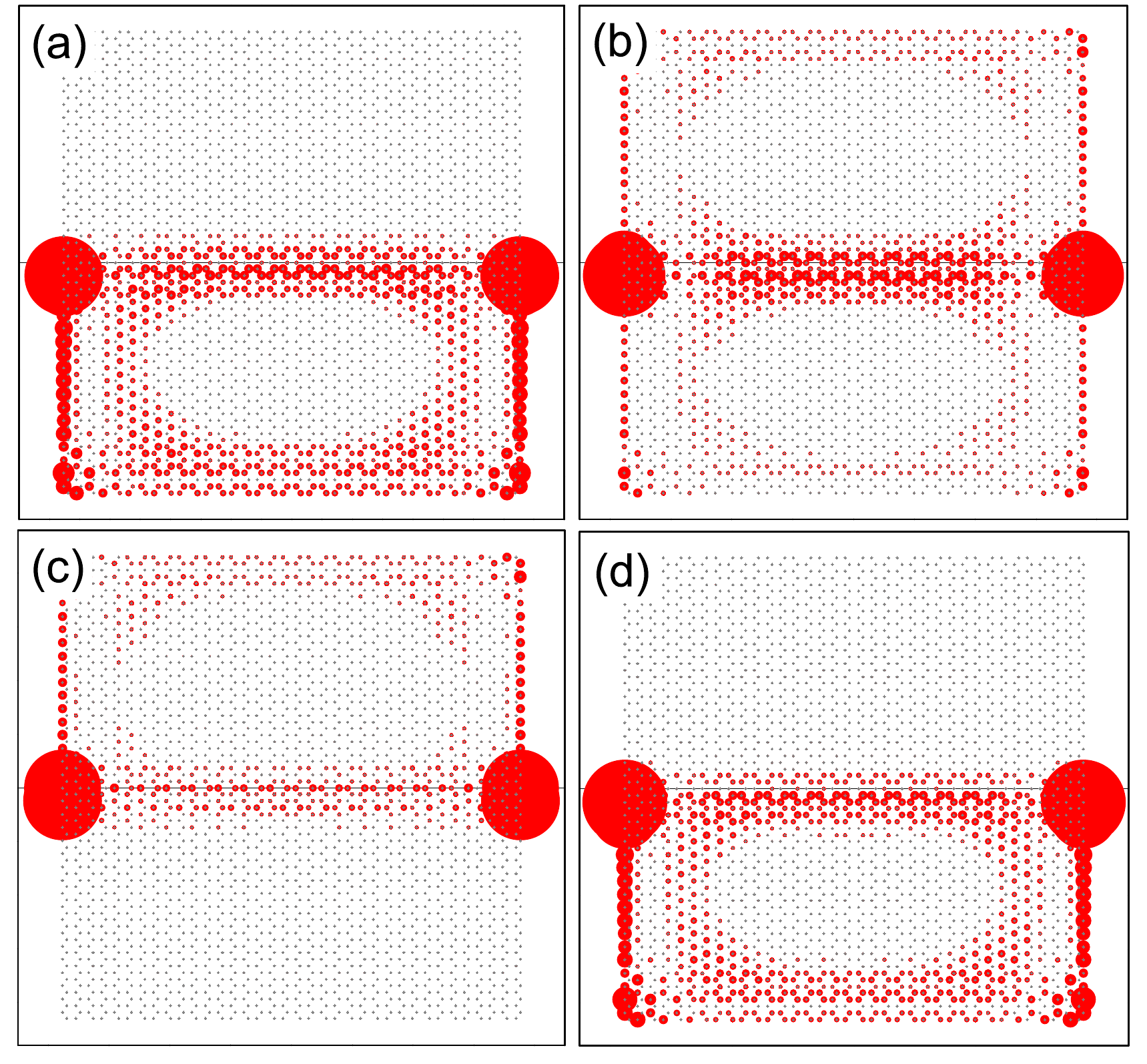}
\caption{(Color online) Probability densities corresponding to the points indicated
by (a-d) in the enlarged circle of Fig. \ref{fig11}(a). The black
line indicates the position of the p-n junction.} \label{fig13}
\end{figure}
\section{snake states: influence of a perpendicular magnetic field}
\subsection{p-n junction}
In the presence of an external magnetic field (see upper panel in
Fig. \ref{fig6} for $\Phi_{c}/\Phi_0=0.1$) the energy spectrum shows
anti-crossings for the energies below the gate voltage amplitude
($E<|U_{b}|$) which is due to the overlap between the quantum Hall
(QH) edge states and the localized states at the p-n interface (i.e.
snake states). Because of the smallness of the dot a large magnetic field (i.e. $B=800$ T for $\Phi_C/\Phi_0=0.1$) is required in order to have a significant influence on the energy levels. Nevertheless, as the influence of the magnetic field scales with the magnetic flux through the dot area, similar results will be obtained for lower magnetic fields if a larger graphene dot is considered. Notice that the number of degenerate levels with
$E=\pm U_{b}$ ($E=0$) in the presence (absence) of a p-n junction does not
change with magnetic field. The red dashed lines in Fig. \ref{fig6}
indicate the Landau levels (LLs) of an infinite graphene sheet that
are shifted up(down) in the presence of an external potential
$U_{b}$($-U_{b}$). The LLs are given by
\begin{equation}
E_{n}=sgn(n)\frac{3at}{2l_{B}}\sqrt{2|n|}\pm U_b
\end{equation}\label{eqLLs}\noindent
where $l_{B}=\sqrt{\hbar/eB}$ is the magnetic length and $n$ is an
integer\cite{zarenia}. Lower panels in Fig. \ref{fig6} show the
probability density corresponding to the states indicated by the arrows in the upper
panel. Panel (a) shows the confinement due to the QH edge states and
zigzag edge states for $U_{b}=0$. In the presence of a p-n junction
and for $E<U_{b}$, states can arise due to the overlap of
the QH edge states and the snake states (see panel (b)) or due to the overlap
with confined states at the p (or n) regions (see panel (d)). For
$E>U_{b}$ the carriers form LLs in the p (or n) potential
regions (see panel (c)).
\begin{figure*}
\hspace{1.5cm}
\includegraphics[width=14 cm]{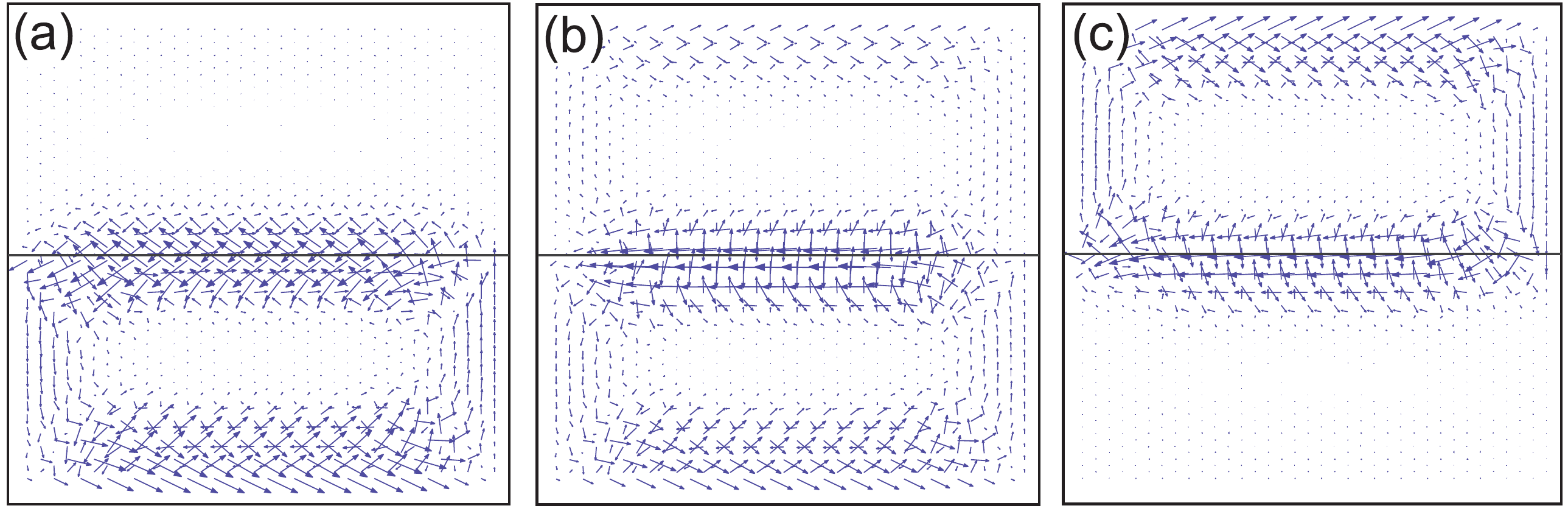}
\caption{(Color online) Current density profile corresponding to the points
indicated by (a-c) in the enlarged circle of Fig. \ref{fig11}(a).
The black horizental line indicates the position of the p-n junction.} \label{fig14}
\end{figure*}
\begin{figure}[!]
\centering
\includegraphics[width=8. cm] {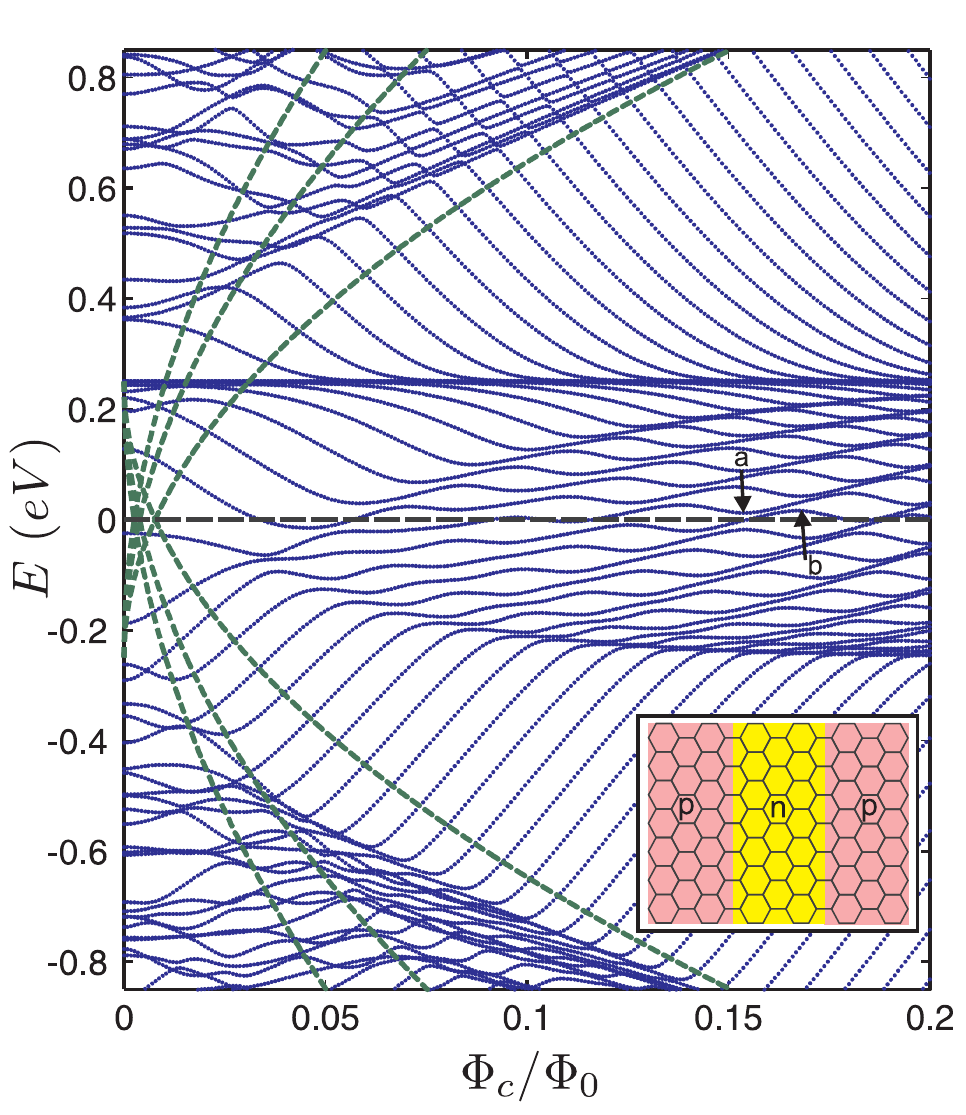}
\caption{(Color online) Energy levels of a rectangular GQD with a p-n-p junction parallel to the zigzag edges as function of external magnetic flux for the same
parameters as Fig. \ref{fig9}. The lower inset illustrates the
system. The number of p-type atoms is twice larger than the number
of n-type atoms. The green dashed curves are the LLs of an infinite
graphene sheet which are shifted up (down) by $U_b$
($-U_b$).}\label{fig15}
\end{figure}

The different types of states become more apparent in 
Fig. \ref{fig7} where we show the current density profile corresponding
to the states shown in the lower panels of Fig. \ref{fig6}. The
current density vectors are obtained using
\begin{equation}
{\boldsymbol j_{l\rightarrow m}}=\frac{i}{\hbar}\big[\langle
\psi_{l}|t_{lm}|\psi_{m}\rangle - \langle
\psi_{m}|t_{lm}^{\ast}|\psi_{l}\rangle \big]
\end{equation}
where ${\boldsymbol j_{l\rightarrow m}}$ is the current flowing out
of site $l$ into site $m$. For clarity we show only the current corresponding to the A sublattice. Figures \ref{fig7}(b,d) clearly demonstrate the presence of snake states at the p-n junction where
we have clockwise and counterclockwise circling currents, respectively, in the $n$
and $p$ regions. The current profile also reflects the direction of the bonds between the carbon atoms and therefore the arrows around the p-n junction sometimes point away from the interface. The streamline plot in the inset of Fig. \ref{fig7}(b) shows the current flow of the snake states more clearly. The vector plot in Fig. \ref{fig7}(a) indicates the cyclotron orbit of a quantum Hall edge state, while Fig. \ref{fig7}(c) shows the current profile of a LL state that is only very weakly influenced
by the p-n junction and the edge of the quantum dot. 

The energy levels as function of $U_b$ are shown in Fig. \ref{fig8},
in the presence of an external magnetic flux $\Phi_{c}/\Phi_0=0.1$
for the dot with p-n junction along the armchair edges (see the
inset of Fig. \ref{fig4}). The red solid lines are the LLs of a
graphene sheet (see Eq. (5)). As in Fig. \ref{fig6}(a) the energy spectrum exhibits different regimes of states;
\textit{i}) The regime of QH edge states where $U_b\leq |E|\leq E_{1}^{LL} $ and $E_{1}^{LL}$ is the first LL obtained from Eq. (5). \textit{ii}) The $|E|\leq U_b$ and $|E|\leq E_{1}^{LL}$ regime where there exist snake states. \textit{iii}) The regime of LLs form for $|E|\geq E_{i\geqslant 1}^{LL}$. Notice that for $U_b>0$ ($U_b<0$) the LLs form at p (n) region. \textit{iv}) The last regime that can be seen in Fig. \ref{fig8} (and Fig. \ref{fig6}(a)) is due to the overlap of the edge states and LLs that occurs when
$E_{i\geqslant1}^{LL}\leq |E|\leq U_b$.      
\begin{figure}[!]
   \subfigure{
   \includegraphics[width=8. cm] {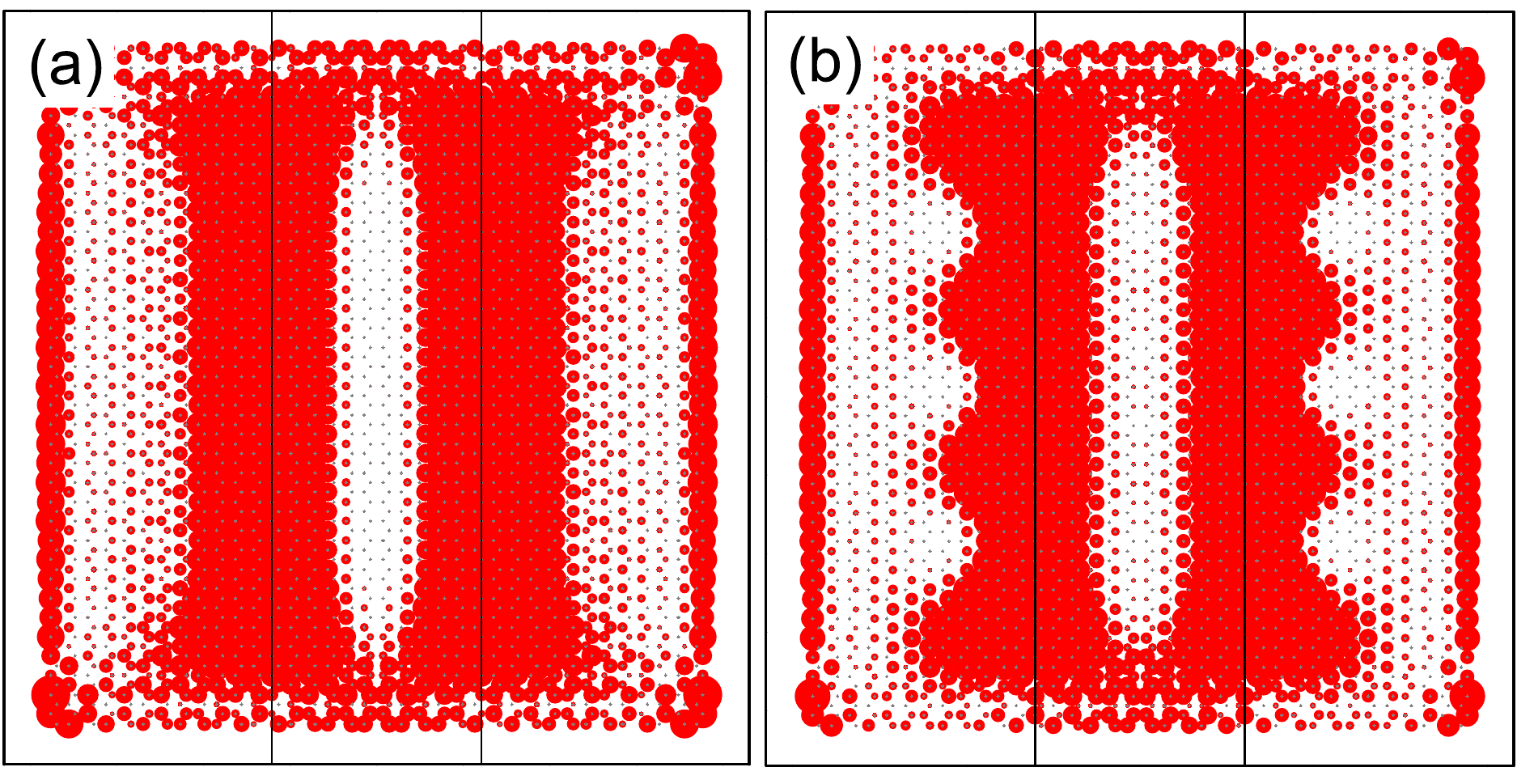}
   }\vspace{0. cm}
    \subfigure{
    \hspace{-0.1 cm}
     \includegraphics[width=8.1 cm] {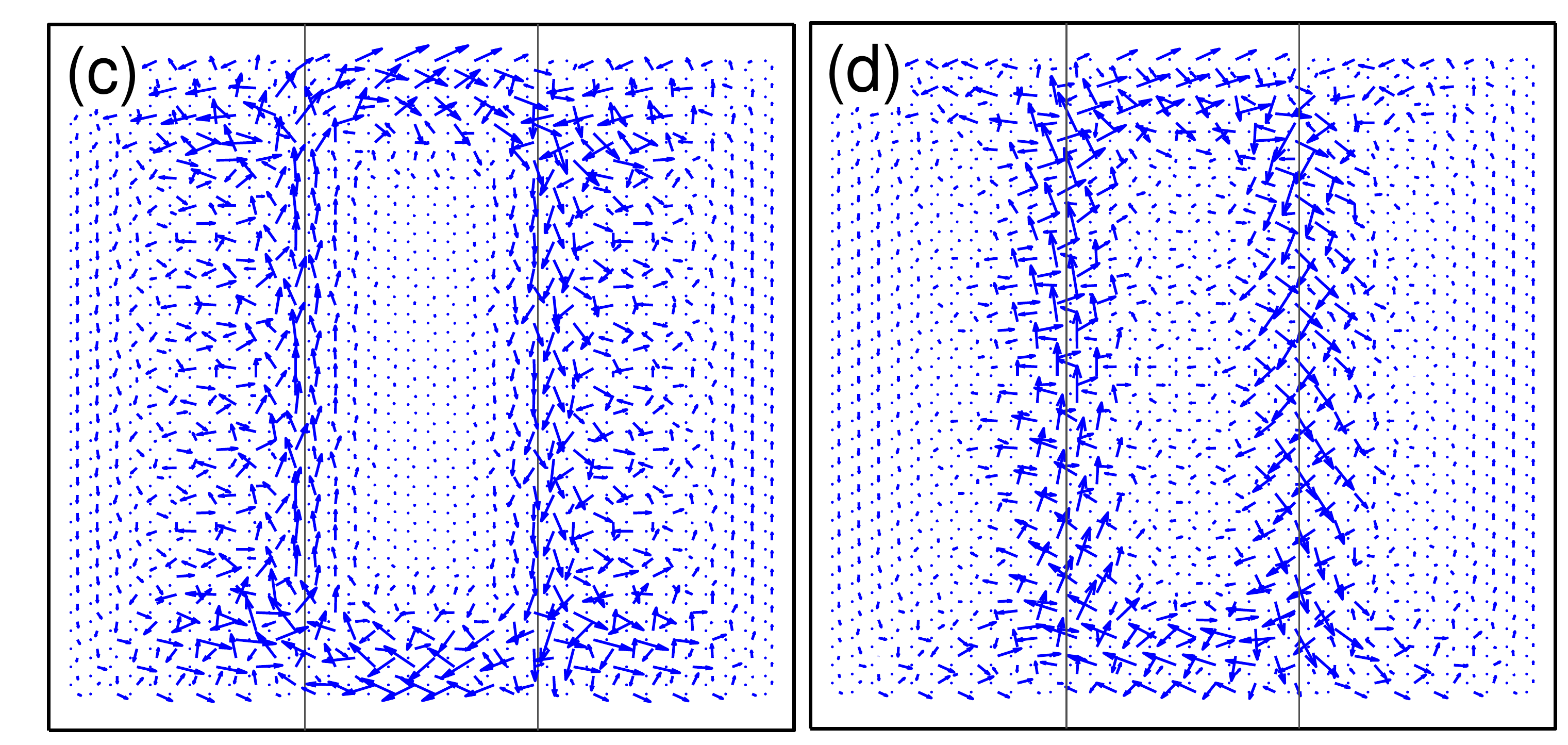}
   }
\caption{(Color online) (a) Electronic probability density and (b) the
corresponding current profile for the points indicated by (a,b) in
Fig. \ref{fig15}. The black vertical lines indicate the position of
the p-n junctions.}\label{fig16}
\end{figure}

Figure \ref{fig9}(a) displays the energy levels of the system
illustrated in the inset of Fig. \ref{fig2} as function of magnetic
flux threading one carbon hexagon $\Phi_{c}$ for $U_b=0.25$ eV and
the same size as Fig. \ref{fig2}. Notice that the zeroth Landau
level in the absence of a gated voltage is now shifted up(down) by
$+U_b$($-U_{b}$). The green dashed curves are LLs of an infinite
graphene sheet (given by Eq. (5)). The magnetic levels in the GQD,
i.e. the so called Fock­ Darwin levels, approach the
LLs\cite{Zhang,Libisch} which are shifted by $\pm U_b$. Some of the
energy levels approach asymptotically the $E=\pm U_{b}$ levels. Due
to the overlap between the QH edge states and the snake states at
the p-n junction, anti-crossings appear in the energy spectrum. An
anti-crossing point around $\Phi_{c}/\Phi_{0}=0.16$ is enlarged in
Fig. \ref{fig9}(a). In Fig. \ref{fig9}(b) the persistent current
corresponding to the first electron (solid curve) and the first hole
(dashed curve) states is shown as function of magnetic flux. The
persistent current $J$ is calculated by taking the derivative of the
corresponding energy levels with respect to the flux as
$J(\Phi_{c})=-\partial E/\partial \Phi_{c}$. Due to the
anti-crossings in the energy spectrum, the persistent current
exhibits an oscillatory behavior with respect to the magnetic flux.
The current oscillation due to the snake states was recently
investigated theoretically for a six-terminal graphene nano-ribbon with a p-n junction\cite{Chen}. Notice that in the presence of the p-n junction the electron QH edge states that are shifted down with $-U_b$ overlap with the hole QH edge states that are shifted up with $U_b$ (they have an opposite circling orbit direction than the electronic QH edge states). This hybridize the states in the region $|E|\leq U_b$ and leads to the current oscillations. Thus, as the magnetic field is adiabatically increased, at each cycle of oscillation the electron becomes predominantly confined either on the p or the n sides of the quantum dot, with the current circulating either clockwise or counterclockwise.  

Figure \ref{fig10} shows the electron probability densities
corresponding to the points indicated by (a-d) in the enlarged
region of Fig. \ref{fig9}(a). Our results indicate that at the
anti-crossing (panel (b)) the carriers are confined by the zigzag
edge atoms and the p-n interface which characterizes the overlap
between the edge and snake states. The points corresponding to the
energy states that increase with respect to the magnetic flux around
the anti-crossing (a,d) are due to states that are confined at the p-n
junction and the zigzag edges in the n-type region (i.e. left side)
while panel (c) displays an electron density that is confined at the right side of the p-n
interface.
\begin{figure}[!]
\centering
\includegraphics[width=8. cm] {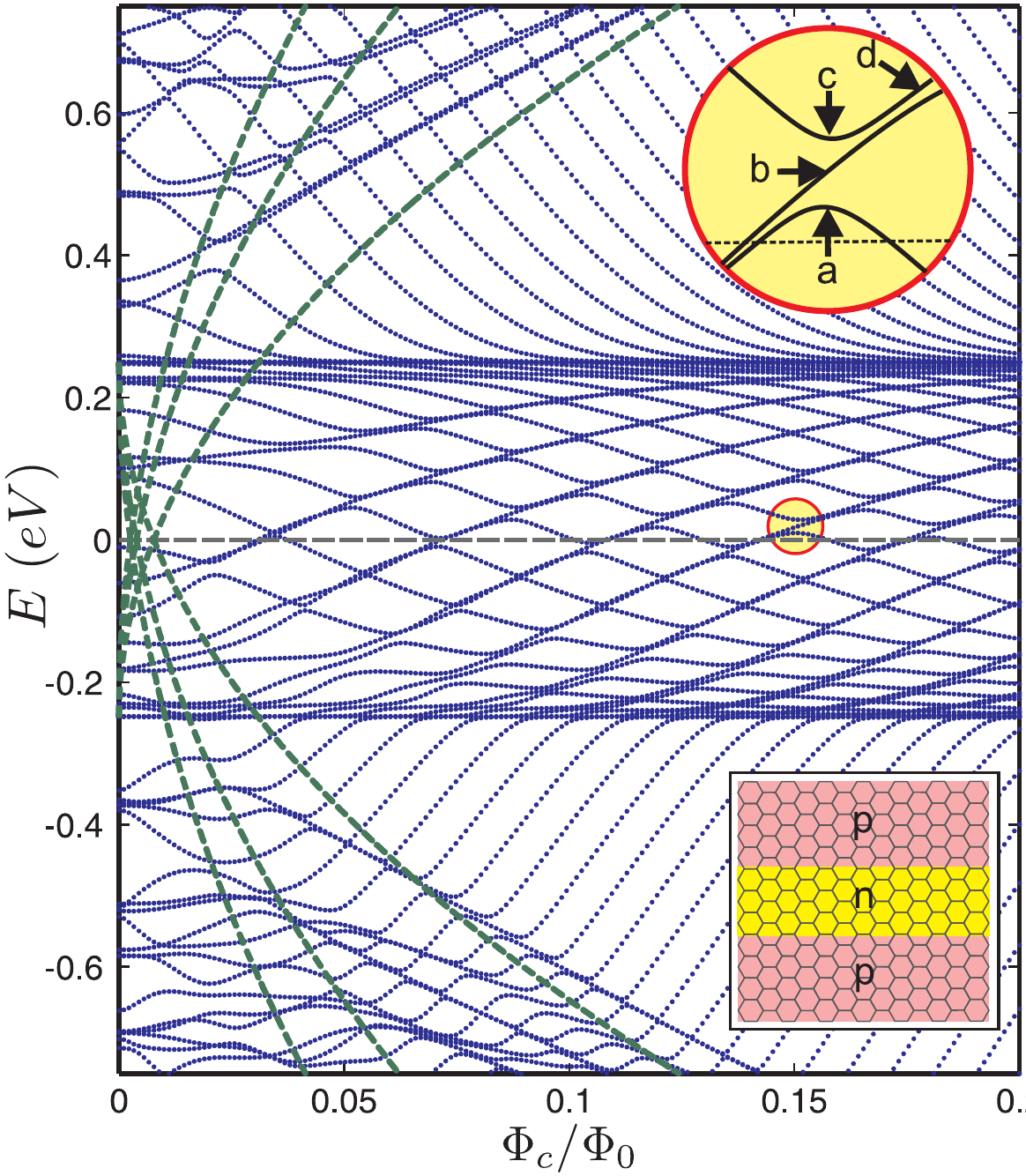}
\caption{(Color online) The same as Fig. \ref{fig15} but for a rectangular GQD with
a p-n-p junction along the armchair edges. The system is illustrated
in the lower inset.}\label{fig17}
\end{figure}

The energy spectrum of the system depicted in the inset of Fig.
\ref{fig4} is shown in Fig. \ref{fig11}(a) as function of magnetic
flux for $U_{b}=0.25$ eV. Since the number of p-type and n-type
atoms are unequal here (where their minimum difference is $N_{a}$)
the energy levels are not symmetric around $E=0$. Now the
confinement due to the p-n junction is along the x-direction which
is perpendicular to the edge states (caused by the zigzag edges).
Therefore the energy spectrum exhibits a distinct behavior from that
of the p-n junction along the zigzag edges. The persistent current
$J$ corresponding to the energy levels indicated by (1) and (2) is
shown in Fig. \ref{fig11}(b) as function of magnetic flux. Notice
that the oscillatory behavior is different from the results in Fig.
\ref{fig9}(b), now the current amplitude decreases smoothly with
increasing magnetic flux. The position of the oscillations are plotted in Fig. \ref{fig12} and compared with the flux through the quantum dot (magenta stars) and half of the quantum dot (blue triangles). Notice that the numerical results are between these two curves. It implies that the effective surface area encircled  by the current is larger than the size of the $n$ or $p$ region. The best fit is obtained for flux through a surface area $S=L_xL_y/1.37$ (see gray squares).
\begin{figure}[!]
\centering\includegraphics[width=8 cm] {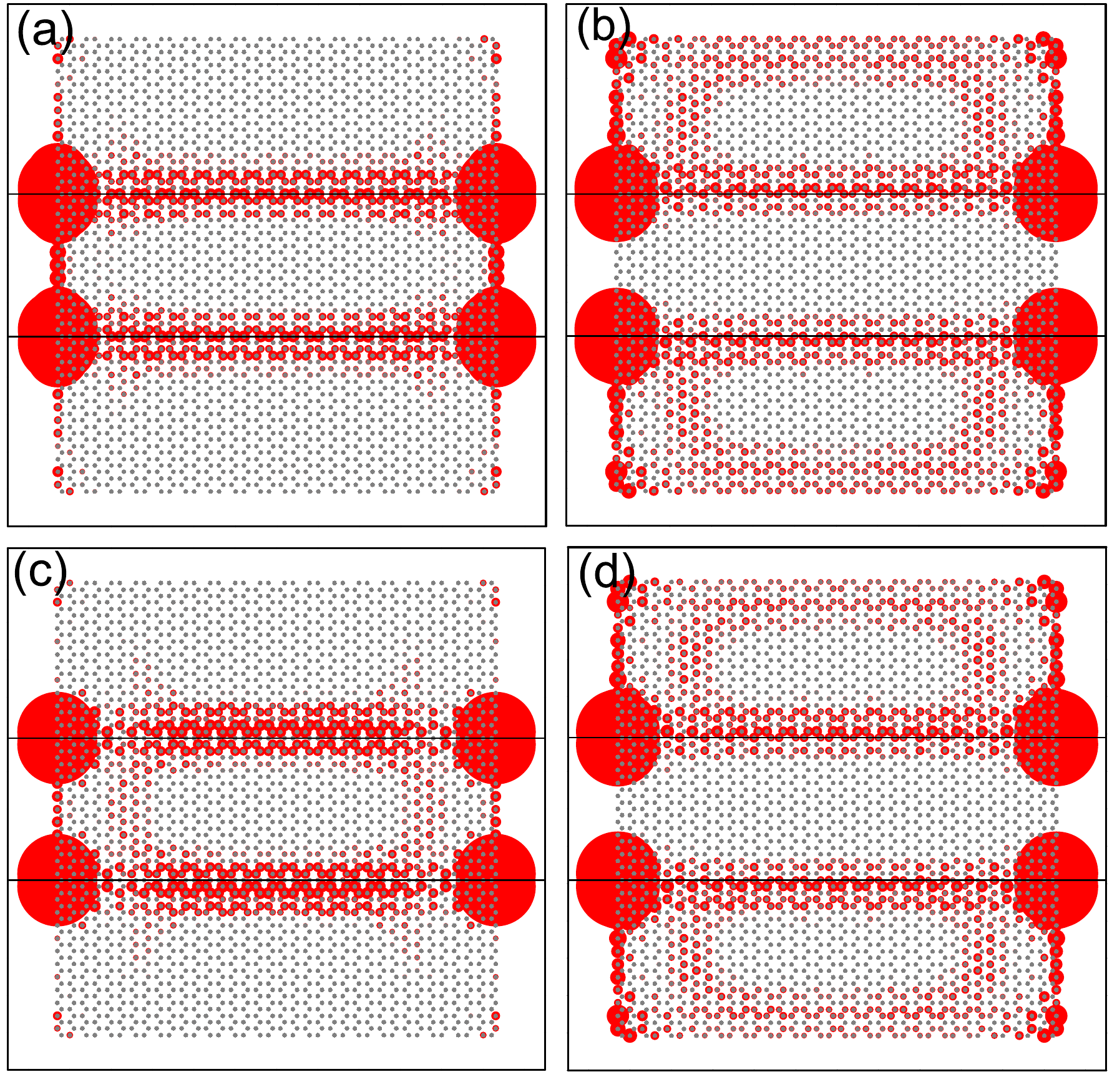}
\caption{(Color online) Probability densities corresponding to the points indicated
by (a-d) in the enlarged region of Fig. \ref{fig17}. The black
horizontal line indicates the position of the p-n
junctions.}\label{fig18}
\end{figure}

Figure \ref{fig13} shows the electron probability densities
corresponding to the indicated points by (a,b,c,d) in the enlarged
circle (an anti-crossing point around $\Phi_{c}/\Phi_{0}=0.16$) of
Fig. \ref{fig11}(a). We have a superposition of three types of
states: \emph{i}) zigzag edge states (modified by the p-n
junction), \emph{ii}) QH edge states where we have skipping orbits, 
and \emph{iii}) snake states. As seen in the figure the overlap of
the confined electron in the snake state and the edge states
leads to a large density at the intersection of the p-n junction and
the zigzag edges. In contrast with the results in Fig. \ref{fig10} only
half of the zigzag edge atoms are contributing to the confinement
due to the edge states. Therefore the carriers are weakly affected
by the edge states in comparison with the confinement due to the snake
states. At the anti-crossing (panel (b)) the electrons are mostly
confined at the p-n interface and along both lengths of the
zigzag edges. The corresponding current profiles of Figs.
\ref{fig13}(a-c) are shown in Fig. \ref{fig14}. Figure 
\ref{fig14}(b) displays the snake states at the p-n interface and
Figs. \ref{fig14}(a,c) show the cyclotron orbit of QH edge states
respectively at the $p$ and $n$ regions.
\begin{figure}[!]
\centering
\hspace{0.5cm}
\includegraphics[width=8 cm] {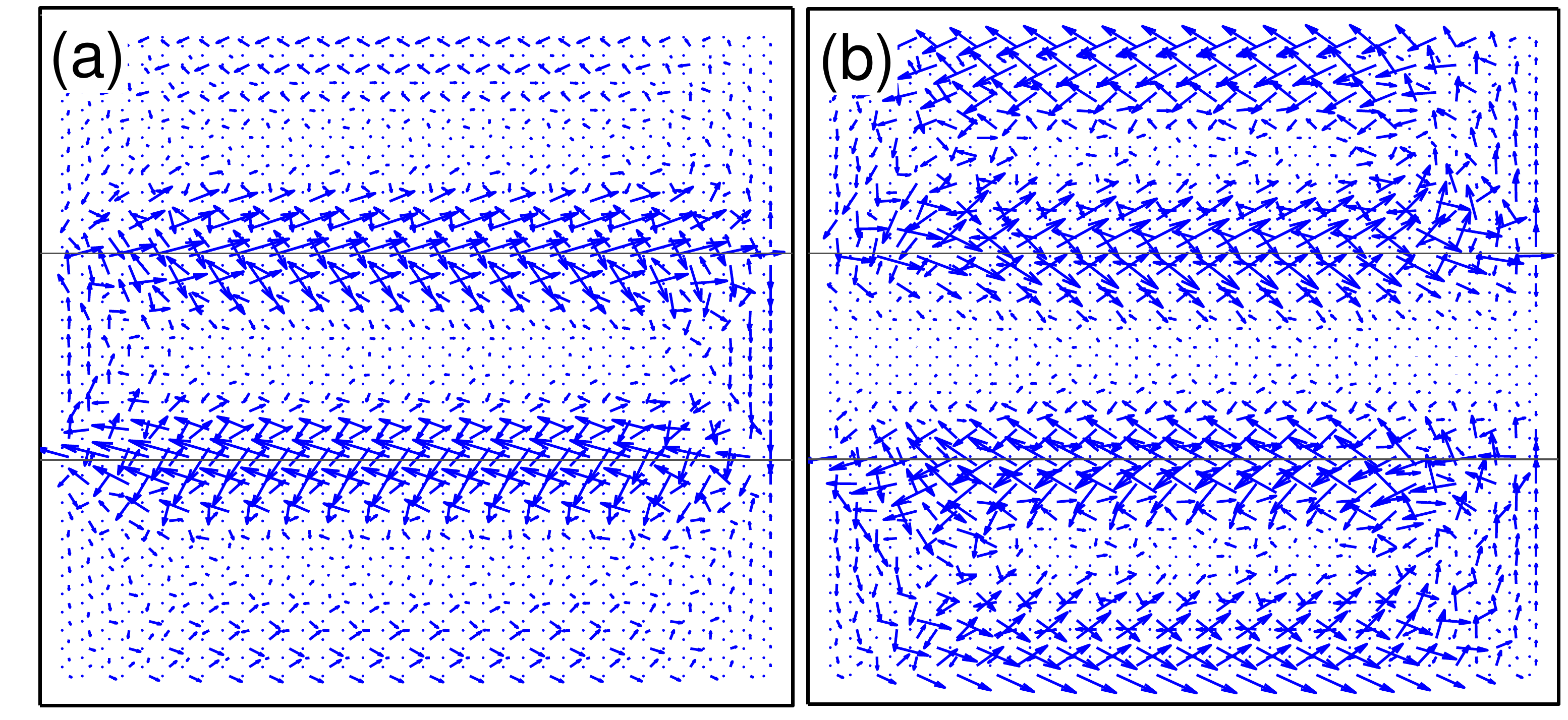}
\caption{(Color online) Current density profile corresponding to the
states shown in Fig. \ref{fig18}(a) and Fig. \ref{fig18}(b). The
black horizontal lines indicate the position of the p-n junctions.}\label{fig19}
\end{figure}
\subsection{p-n-p junction}
Next we investigate the effect of multiple p-n junctions where we
limit ourselves to the example of two junctions. We want to know if there can
be any interplay between the two junctions, i.e. can states 
be confined over the two junctions? Will there be circling currents
between the two junctions?
\begin{figure}[!]
\centering\
   \includegraphics[width=8. cm] {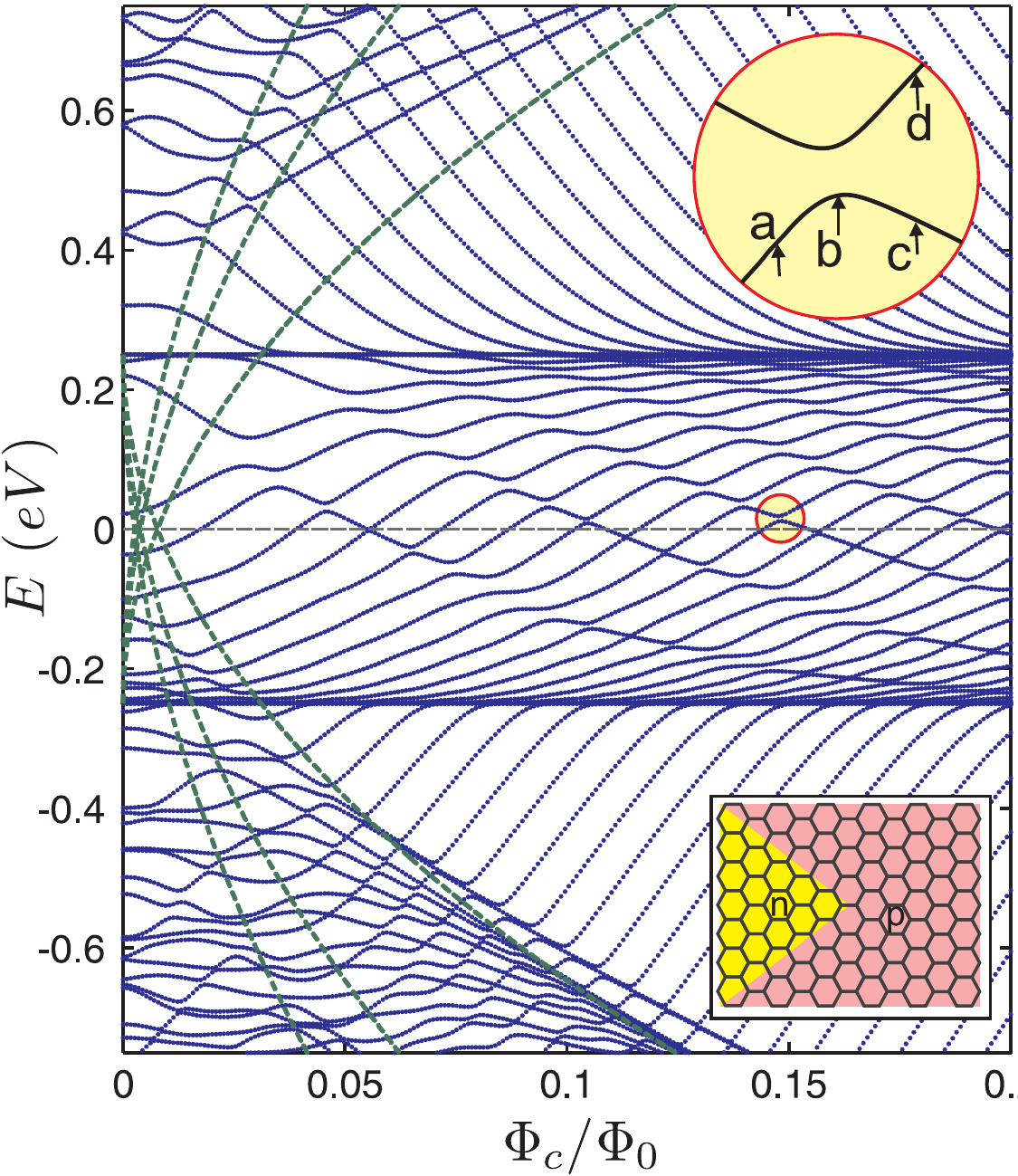}
\caption{(Color online) Energy levels of a rectangular GQD with triangular shaped
p-n junction as function of external magnetic flux for the same
parameters as Fig. \ref{fig9}. The lower inset shows schematically the system. 
The green dashed curves are the LLs of an infinite
graphene sheet which are shifted up (down) by $U_b$
($-U_b$).}\label{fig20}
\end{figure}
\begin{figure}[!]
\centering \includegraphics[width=8 cm] {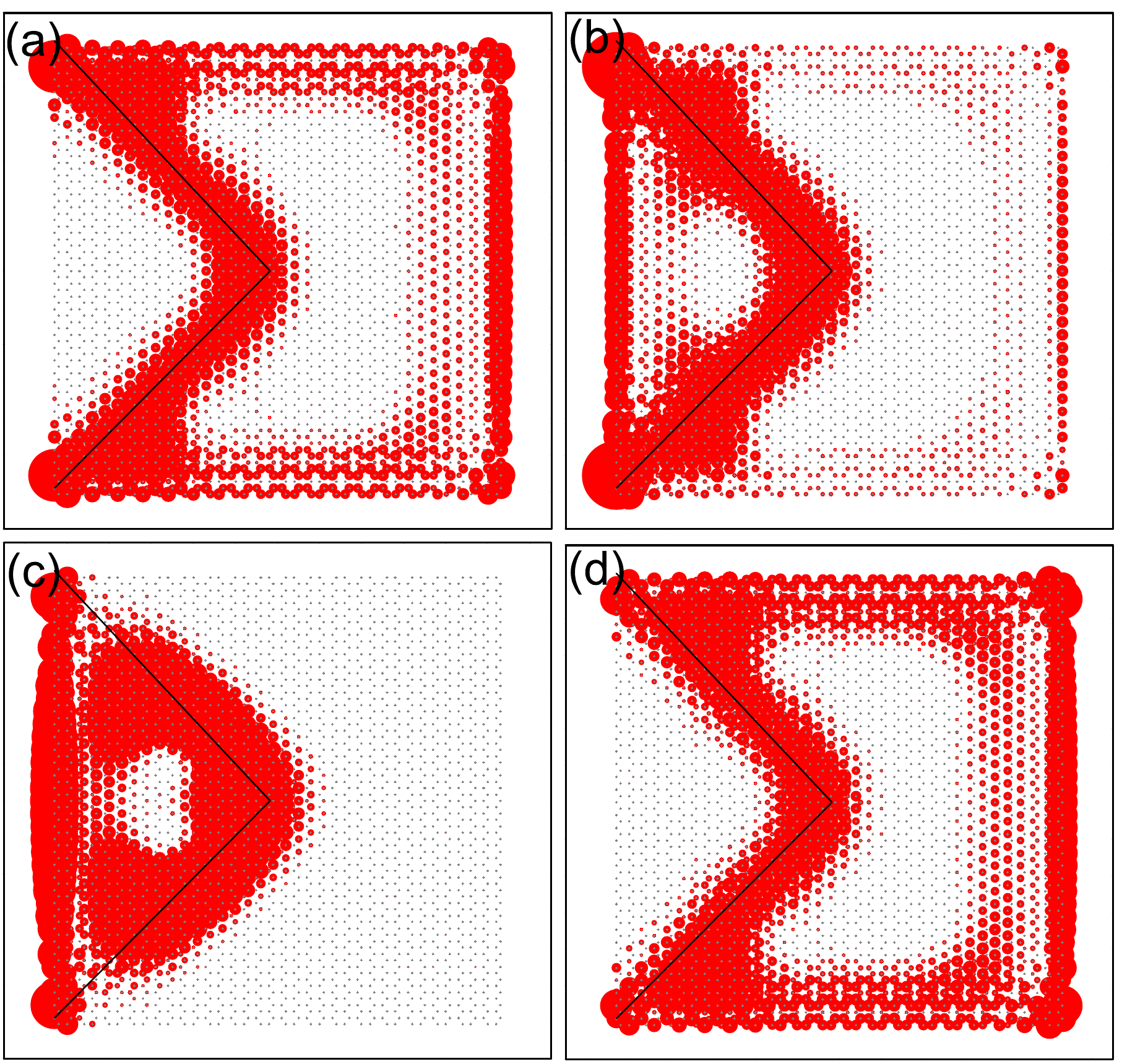}
\caption{(Color online) Probability densities corresponding to the points indicated
by (a-d) in the enlarged region of Fig. \ref{fig20}. The black line
indicates the position of the p-n junction.}\label{fig21}
\end{figure}
\begin{figure}
\centering
\hspace{-1cm}
\includegraphics[width=8.5 cm] {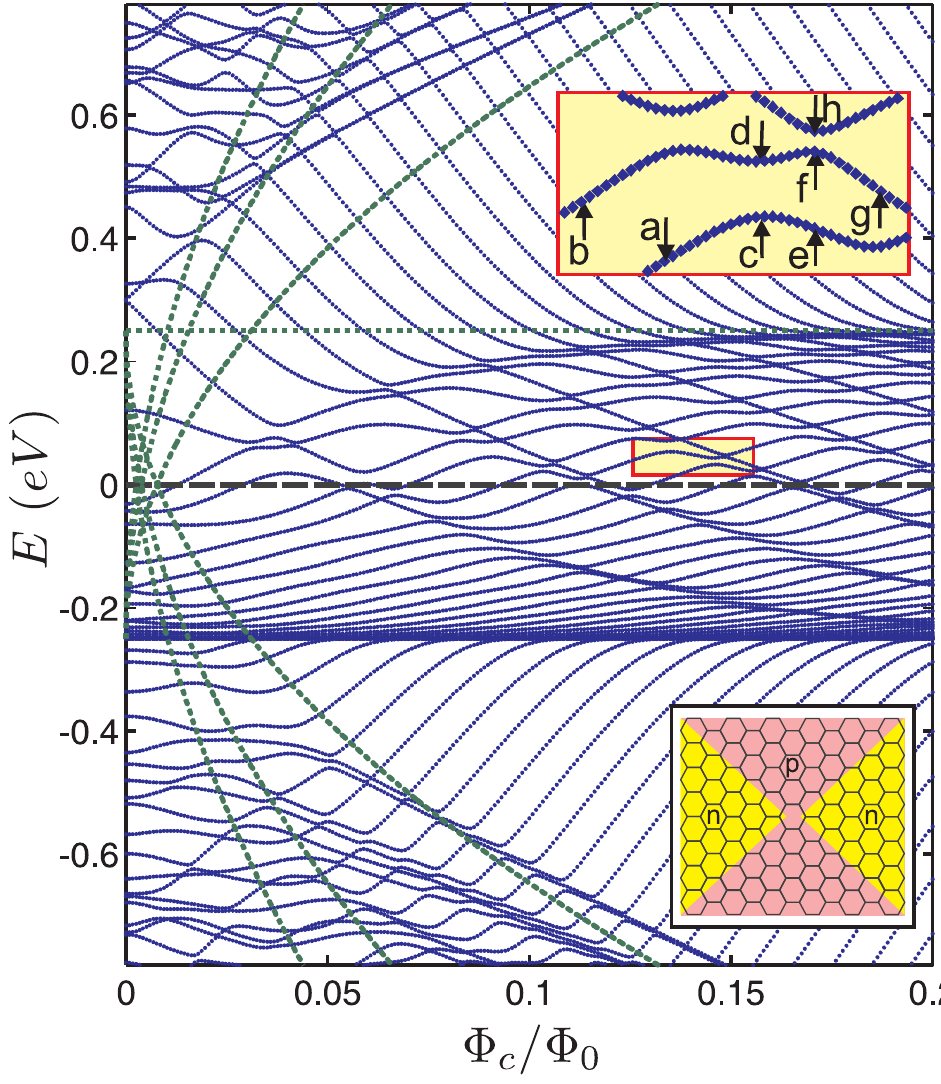}
\caption{(Color online) Energy levels of a rectangular GQD with triangular shaped
p-n-p junction as function of external magnetic flux for the same
parameters as Fig. \ref{fig9}. The lower inset shows the system
schematically where the yellow regions with zigzag edges indicate
the n-type atoms. The green dashed curves are the LLs which are
shifted up (down) by $U_b$ ($-U_b$).} \label{fig22}
\end{figure}

A schematic illustration of a p-n-p junction in a rectangular GQD is
depicted in the inset of Fig. \ref{fig15} where the p-n-p junction
is parallel to the zigzag edges (along y-direction). The
corresponding spectrum in Fig. \ref{fig15} exhibits quite distinct
anti-crossings from the case of the p-n junction. On the other hand
in low magnetic fields the hole edge states (i.e. those hole states
that decrease with respect to the magnetic flux) do not approach the
zeroth Landau Level ($E=-U_{b}$) which is a consequence of the fact
that the n-type region does not have a boundary with the zigzag
edges. Notice that the hole edge states approach $E=-U_b$ in high
magnetic fields. The electron probability densities for the points
indicated by (a) and (b) are shown in Figs. \ref{fig16}(a,b) where
the densities are spread out mostly along the p-n and n-p interfaces. The corresponding current
profiles are plotted in Figs. \ref{fig16}(c,d). Our results show opposite circling currents between the two junctions.
\begin{figure*}[]
\begin{center}
\includegraphics[width=16cm]{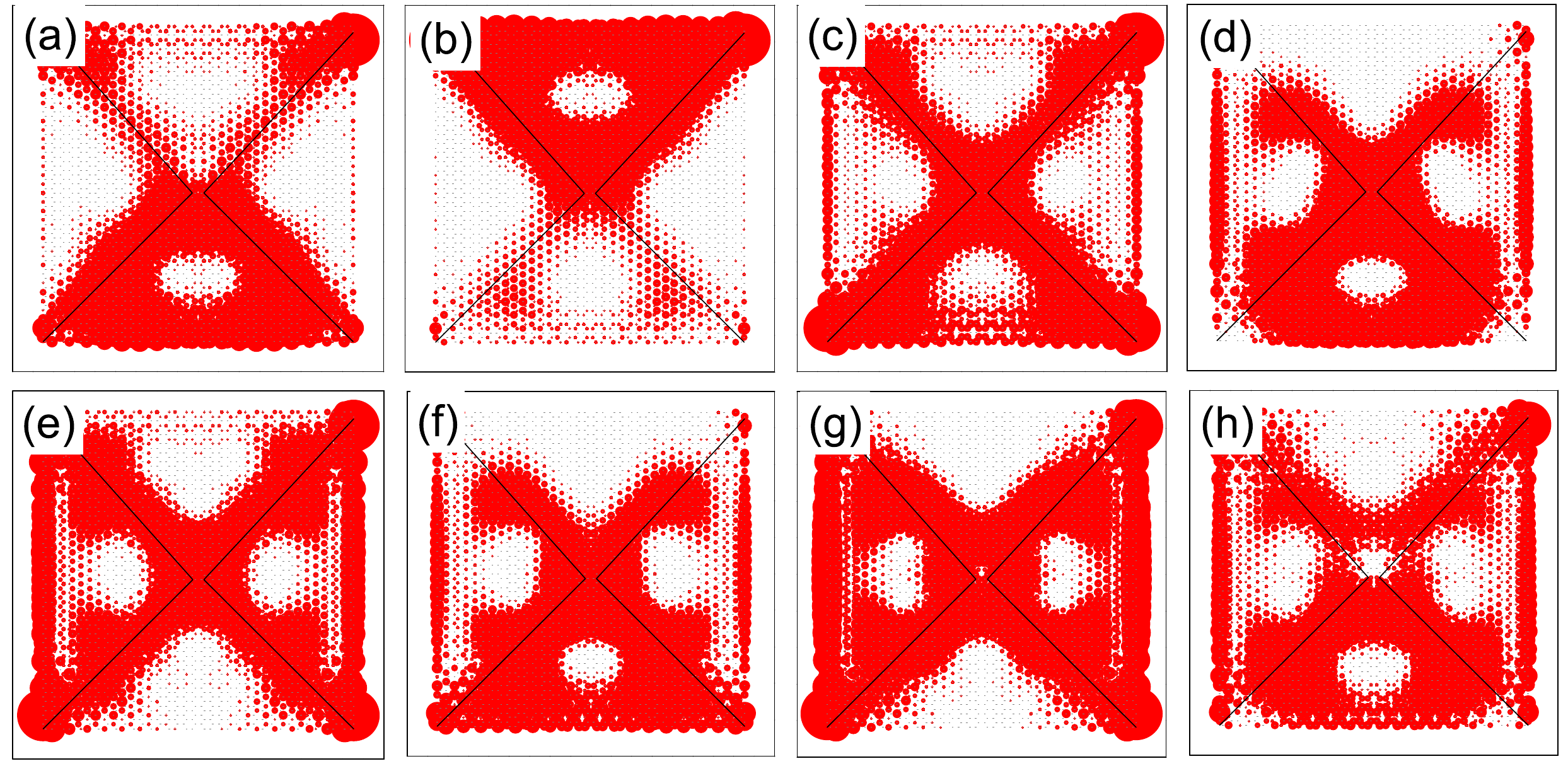}
\caption{(Color online) Probability densities corresponding to the points indicated
by (a-h) in the enlarged region of Fig. \ref{fig22}. The black lines
indicate the position of p-n junctions.} \label{fig23}
\end{center}
\end{figure*}

Figure \ref{fig17} displays the energy spectrum for the p-n-p
junction parallel to the armchair edges in a rectangular GQD. The
system is depicted in the lower inset. Here the n-type region is
connected to the zigzag lengths which leads to the convergence of
the hole edge states to $E=-U_{b}$ even in low magnetic fields. As
in previous cases for the region $|E|<U_{b}$ anti-crossings appear in
the spectrum. The electron probability densities corresponding to
the points around the enlarged anti-crossing (yellow circle) are shown in Fig. \ref{fig18}. Note that the anti-crossing behaviour
is qualitative distinct from the previous cases shown in 
Figs. \ref{fig9}(a) and \ref{fig11}(a). Our results indicate that at the
anti-crossing (a,c) the electrons are confined due to the edge
states corresponding to the zigzag atoms in the n-type region and
the snake states. The points (b,d), correspond to states that ate confined at
the zigzag edges in the p-type regions and near the p-n-p junction. In all points, the probability density shows strong peaks at the intersection of the p-n junction with the zigzag edges. The density profiles have a dumb-bell shape. As before, these localized states are associated with the hybridization of the zigzag edge states on each side of the junction. However, for non-zero magnetic field these states overlap with snake states that propagate along the potential interfaces. The current profile corresponding to Figs. \ref{fig18}(a,b) is shown in Fig. \ref{fig19}(a,b) where the counter-circling cyclotron orbits in
the n and p regions demonstrate the existence of snake states at the
p-n interfaces.
\vspace{-0.4 cm}
\section{Triangular shaped p-n junction}
Next we consider the effect of the gate shape on the energy
spectrum. Figure \ref{fig20} displays the energy levels for the
system illustrated in the lower inset where a triangle-shaped gate
voltage is assumed for the n-type region. Notice that here we choose
an arbitrary direction for the p-n junction and it not necessarily
matched with the zigzag or armchair direction. Since the number of
n-type and p-type atoms are unequal the electron and hole energy
levels are not symmetric, i.e. $|E_{e}(\Phi_{c})|\neq
|E_{h}(\Phi_{c})|$. Due to the confinement by both edge states (at
zigzag edges) and snake states (at p-n interface) anti-crossings
appear in the energy spectrum. An enlargement around one of the
anti-crossings at $\Phi_{c}/\Phi_{0}=0.15$ is shown in the inset of
Fig. \ref{fig11}(b). The electron probability densities for the
points around this anti-crossing (a,b,c,d) are shown in Fig.
\ref{fig21}. Panel (b) shows the electron density at the
anti-crossing where the electrons confined along the
zigzag edge and the p-n interface. For the points whose energy increases with flux (a,d) the electron is localized along the zigzag edge in the $p$ region and snake states are present along the p-n junction while the probability density for the point (c) is mostly along the zigzag edges in the n region and the p-n
interface.

As a last example we investigate the energy spectrum of a system consisting of a
point contact in a rectangular GQD (see the lower inset of Fig.
\ref{fig22}). Recently, transport measurements of such a system were
carried out, and it was found that, due to a chaotic mixing of edge
channels an unexpected half-integer plateau was observed in the
QH resistivity\cite{Marcus2}. The spectrum exhibits double
anti-crossings between the energy levels in the region $|E|<U_{b}$.
The enlarged rectangle in Fig. \ref{fig22} shows one of these
double-anticrossings around $\Phi_{c}/\Phi_{0}\approx0.15$. Figure
\ref{fig23} shows the electron probability densities corresponding
to the points indicated by a-h in Fig. \ref{fig22}. The energy
levels between $-U_b$ and $U_b$ that increase with magnetic flux are
due to the overlap of the QH edge states in the p region and the
snake states at the p-n interface (Figs. \ref{fig23}(a,b)). Those
energy levels that decrease with magnetic flux correspond to the zigzag
edge states which hybridize with the QH edge states in the n-region
and to snake states (Figs. \ref{fig23}(e,g)). Notice that at the
anti-crossings, see Figs. \ref{fig23}(c,d,f,h), we have an overlap
of three types of localized states (i.e. QH edge states, snake states and
zigzag edge states).
\section{Concluding remarks}
We presented numerical results for the energy spectrum and magnetic
field dependence of the eigenstates of graphene-based quantum dots,
on which p-n junctions create electron and hole-doped regions. The
presence of the magnetic field, together with the coupling between
electron and hole states across the potential barrier due to Klein
tunneling leads to the appearance of localized states at the
potential interface, known as snake states. These states, which have
previously been investigated for pn junctions on infinite graphene
sheets, can influence the transport properties of graphene-based
nanodevices. We have obtained results that show that for the case of
quantum dots the low energy dynamics of the system is dominated by
hybridized states that arise due to the overlap between quantum Hall
edge states and the snake states at the p-n junction, with the snake
states allowing the superposition of quantum Hall edge states at the
p and n sides of the dot. These states are characterized by an
energy spectrum that displays an oscillating behavior as function of
the electrostatic potential and magnetic field at the vicinity of
the Fermi energy. Furthermore, the energy spectrum was shown to
depend on the specific alignment of the potential interfaces with
regard to the graphene lattice, as well as on the geometry of the
gates. The dots were assumed to be defect-free and to have perfect
zigzag or armchair edges. Future work shall concentrate on the
effect of edge disorder, impurities and defects on the electronic properties of
these structures. Another aspect that shall be considered is the
influence of the particular choice of the potential profile and shape of
the graphene flake on the confined states. 
\section{Acknowledgment}
This work was supported by the Flemish Science Foundation (FWO-Vl),
the European Science Foundation (ESF) under the EUROCORES program
EuroGRAPHENE (project CONGRAN), the Brazilian agency CNPq (Pronex),
and the bilateral projects between Flanders and Brazil and the
collaboration project FWO-CNPq.

\end{document}